\documentclass[%
 reprint,
 amsmath,amssymb,
 aps,
 noeprint,
 prx,
]{revtex4-2}

\usepackage{graphicx}
\usepackage{xcolor}
\usepackage{dcolumn}
\usepackage[ruled,vlined]{algorithm2e}
\usepackage{braket}
\usepackage{mathtools}
\usepackage{bm}
\usepackage[colorlinks,linkcolor=blue,citecolor=red,urlcolor=blue]{hyperref}
\usepackage{times}

\usepackage{tikz}
\newcommand*{\tikzcircle}[1]{\begin{tikzpicture}[scale=0.1]%
    \draw (0,-2) circle (0.8);
    \fill[fill opacity=1.0,fill=black] (0,-2) -- (90:-1) arc (90:90-#1*3.6:1) -- cycle;
    \end{tikzpicture}}

\newcommand{\ecircle}{\tikzcircle{0}}
\newcommand{\fcircle}{\tikzcircle{100}}

\newcommand{\nn}{\nonumber}

\newcommand{\abs}[1]{\lvert#1\rvert}

\newcommand{\tr}{\mathrm{Tr}}
\newcommand{\pf}{\mathrm{Pf}}

\DeclarePairedDelimiter{\norm}{\lVert}{\rVert}

\newcommand{\HarvardPhysics}{Department of Physics, Harvard University, Cambridge, MA 02138, USA}

\begin{document}

\title{Non-Abelian Floquet Spin Liquids in a Digital Rydberg Simulator}

\author{Marcin Kalinowski}
\affiliation{\HarvardPhysics}

\author{Nishad Maskara}
\affiliation{\HarvardPhysics}

\author{Mikhail D. Lukin}
\affiliation{\HarvardPhysics}

 \date{\today}

\begin{abstract}
Understanding topological matter is an outstanding challenge across several disciplines of physical science.  Programmable quantum simulators have emerged as a powerful approach to studying such systems. 
While quantum spin liquids of paradigmatic toric code type have recently been realized in the laboratory,  controlled exploration of topological phases with non-abelian excitations remains an open problem. We introduce and analyze a new approach to simulating topological matter based on periodic driving. Specifically, we describe a model for a so-called Floquet spin liquid, obtained through a periodic sequence of parallel quantum gate operations that effectively simulates the Hamiltonian of the non-abelian spin liquid in Kitaev's honeycomb model. We show that this approach, including the toolbox for preparation, control, and readout of topological states, can be efficiently implemented in state-of-the-art experimental platforms. One specific implementation scheme is based on Rydberg atom arrays and utilizes recently demonstrated coherent qubit transport combined with controlled-phase gate operations. 
We describe methods for probing the non-abelian excitations, and the associated Majorana zero modes, and simulate possible fusion and braiding experiments. Our analysis demonstrates the potential of programmable quantum simulators for exploring topological phases of matter. Extensions including simulation of Kitaev materials and lattice gauge theories are also discussed.
\end{abstract}

\maketitle

\section{Introduction}

Techniques for quantum simulation of topological quantum matter and lattice gauge theories are now being actively explored. Since many of these systems are difficult to treat analytically or to simulate on classical computers, often even fundamental concepts are not well understood.
Of particular interest are topological phases~\cite{Anderson}, whose long-range entanglement generates a host of interesting properties, such as emergent gauge fields~\cite{Wilczek,Sachdevusv}, quantum error correcting codes~\cite{Kitaev.2003}, and excitations with non-trivial statistics~\cite{Nayak.2008}.
Recent developments in superconducting qubit technology~\cite{Google.2021} and neutral atom arrays~\cite{Ebadi.2021,Bluvstein.2022,Scholl.2021} have provided exciting signatures of $\mathbb{Z}_2$ topological order~\cite{Semeghini.2021,Google.2021}, the simplest realization of an abelian topological order. 
An outstanding challenge is developing methods which would enable these platforms to explore and control \emph{non-abelian} topological order. In such systems, the exchange of quasi-particles can result in non-abelian unitary operations acting on a degenerate set of ground states, which encode protected quantum information (logical qubits). This property makes the non-abelian anyons not only a fascinating condensed-matter phenomenon but also a building block of topological quantum computation (TQC)~\cite{Kitaev.2003,Nayak.2008}, since the logical subspace is robust against local errors and protected by a gap to the rest of the spectrum. While there has been a long-lasting effort to realize this kind of topological order in solid-state systems~\cite{Takagi.2019} and several theoretical proposals have been put forward to probe it with synthetic  systems, such as cold atoms~\cite{Duan.2002,Liu.2013,Sun.2022} and polar molecules~\cite{Micheli.2006,Manmana.2013}, the direct observation of its exceptional properties has thus far eluded experimental realization.

In this work, we introduce an approach to creating and controlling non-abelian topological matter based on periodic modulation.
Specifically, we show how the gapped non-abelian phase of the Kitaev honeycomb model can be effectively realized as a \emph{Floquet spin liquid}, generated by time evolution under a repeating sequence of  two-body Ising Hamiltonians~\cite{Bookatz.2014}.
In particular, while the time-averaged Hamiltonian corresponds to the gapless phase in Kitaev's model, the first-order correction breaks time-reversal symmetry and induces a finite energy gap, which is crucial for stabilizing the non-abelian excitations.

We demonstrate that such a Floquet spin liquid can be implemented in a hardware-efficient way using programmable arrays of neutral atoms. This approach combines coherent qubit transport with parallel two-qubit controlled phase gates and global single-qubit rotations to realize the Floquet spin liquid in a digital fashion.
Using these efficient primitives, we develop a Hamiltonian simulation toolkit to prepare, control, and measure topological matter. By combining physical insights and variational optimization, we first demonstrate how to reliably perform time evolution and prepare ground states of the honeycomb model in the non-abelian phase.
Subsequently, we show that this toolbox can be used to study the defining properties of the chiral non-abelian theory at different levels of experimental complexity: The chiral nature of the edge modes can be probed by performing a simple quench of the boundary conditions while a minimal additional overhead allows us to create and adiabatically transport the non-abelian Majorana particles characteristic of the Kitaev $B$ phase.

These Majorana zero modes are fractionalized fermions and when brought together can fuse to either an occupied or unoccupied fermion mode.
We show how the tools developed here can be used to implement braiding and fusion experiments, probing their non-abelian nature.
In particular, the degenerate subspace formed by multiple Majoranas can be manipulated by exchanging particles which is the basis for topological quantum computing. 
We furthermore design a dynamical protocol to readout the local fermion content---a necessary component for characterizing the fusion and braiding rules of the anyonic theory. By introducing a local magnetic field, we couple the fermion to a system with $\mathbb{Z}_2$ vortices, which correspond to simple qubit observables. 
Then, we leverage an emergent two-level system analogy to design a composite pulse sequence which improves the fidelity of the particle-to-vortex mapping.

We note that the Majorana modes can also be created in the abelian toric code phase, either 
at the
intersection of $e$ and $m$ boundaries, or in the bulk at the endpoints of 1D lattice defects~\cite{Bombin,Yuri}. Recent experiments~\cite{Andersen} explored gate-based manipulation of such states without energetic protection provided by the spectral gap. In contrast, the Floquet spin-liquid phase and the non-abelian anyons 
explored in this work are analogous, respectively, to $p+ip$ superconductors and the corresponding  Majorana modes pinned to vortex excitations. 
The anyons discussed in this work are pinned to \textit{pseudo}-vortices, which reside at the ends of flipped-bond strings.

Our method can also be adapted to implement generalizations of the Kitaev honeycomb model with long-range and many-body interaction, or external fields. These systems are subjects of extensive theoretical and experimental studies~\cite{Takagi.2019}, and cannot be simulated efficiently on a classical device, making them prime candidates for quantum simulation.
Finally, the tools developed here for the implementations of generalized Kitaev models can also be used to study integrability-breaking quenches and to potentially simulate lattice gauge theories in (1+1)D with substantially shorter circuit depths compared to existing methods.

\section{Floquet spin liquid}
\begin{figure*}
    \centering
    \includegraphics[width=\linewidth]{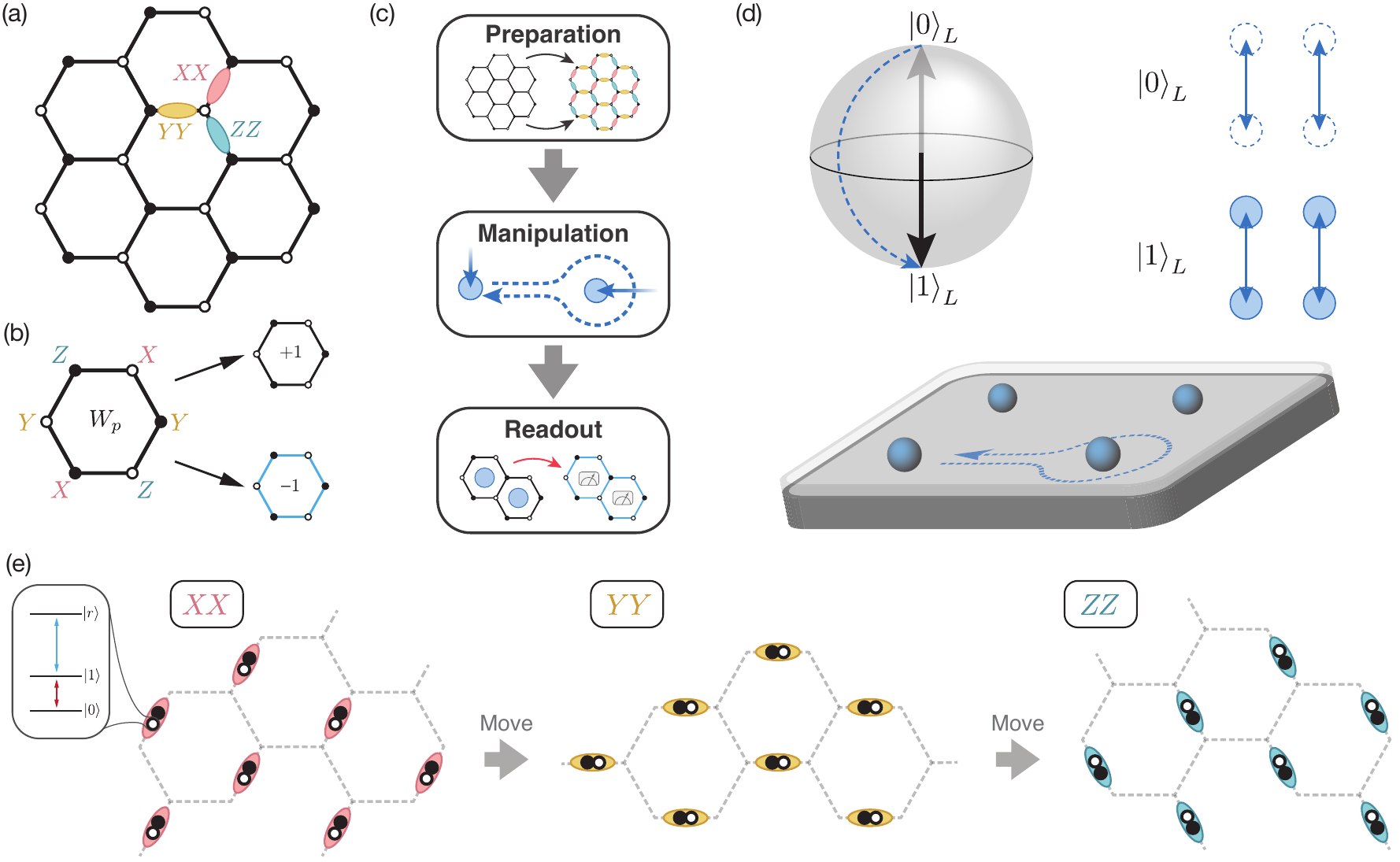}
    \caption{\textbf{Kitaev honeycomb model on a digital Rydberg simulator.} (a) The three types of spin-spin interactions $H_X$, $H_Y$, and $H_Z$ applied along the three directions of the honeycomb lattice together form the highly anisotropic Kitaev honeycomb model described by the Hamiltonian~$H_0$ in Eq.~\eqref{eq:H0}. After breaking the time-reversal symmetry, this model hosts a gapped non-abelian phase. (b) The plaquette operators $W_p = Y_1 Z_2 X_3 Y_4 Z_5 X_6 = \pm 1$ commute with all the link operators $H_X,H_Y,H_Z$ and form an extensive set of conserved quantities.  We associate the presence of a $\mathbb{Z}_2$ vortex at the plaquette $p$ if $W_p=-1$. The ground state has no vortices, i.e., $W_p=+1$ for all $p$. (c) In this work we develop the toolbox for the three capabilities necessary to fully probe a non-abelian system on a digital simulator: state preparation, manipulation, and readout. (d) Four Majorana zero modes encode a topological qubit, as they span a two dimensional Hilbert space (for the case shown here) that is separated by the gap from the rest of the spectrum. The logical states $\ket{0}_L,\ket{1}_L$ correspond to the unoccupied and occupied zero mode, respectively. Measurement in the logical $Z_L$ or $X_L$ basis is performed by fusing the majoranas in an appropriate configuration, and measuring the presence of a fermion particle. Moving the Majorana particles around each other performs non-trivial unitary operations in the logical subspace---a manifestation of their non-abelian nature.  (e) The spin degrees of freedom are encoded in the magnetically insensitive hyperfine states $\{\ket{0},\ket{1}\}$ of $^{87}$Rb atoms and a highly excited Rydberg state $\ket{r}$ is used to perform two-qubit entangling gates. We utilize reconfigurable Rydberg arrays to implement the Floquet Hamiltonian of the Kitaev spin liquid in a parallel fashion. This approach is simple, scalable, and all of its components have already been demonstrated in recent experiments~\cite{Bluvstein.2022}.}
    \label{fig:1}
\end{figure*}

First, we describe the dynamical mechanism for creating the gapped non-abelian phase in a periodically driven system.
Consider a two-dimensional honeycomb lattice with three types of interactions in the three different directions~[Fig.~\ref{fig:1}(a)]. 
Their respective Hamiltonians are,
\begin{subequations}
\begin{align}
    H_X &= -J_X \sum_{\braket{i,j}_X} X_i X_j,\label{eq:Hx} \\
    H_Y &= -J_Y \sum_{\braket{i,j}_Y} Y_i Y_j,\label{eq:Hy} \\
    H_Z &= -J_Z \sum_{\braket{i,j}_Z} Z_i Z_j,\label{eq:Hz}
\end{align}
\end{subequations}
where $\braket{i,j}_{X/Y/Z}$ denotes the appropriate set of links. Each of these two-body interactions can occur naturally in a magnetic system, but due to their extreme spatial anisotropy it is difficult to realize the three of them, $H_0 = H_X+H_Y+H_Z$, at the same time.

In a digital simulation, however, it is possible to periodically apply the interactions \eqref{eq:Hx}-\eqref{eq:Hz} one at a time. Then, over one driving period, the unitary evolution is described by
\begin{equation}\label{eq:Floquet}
    U(\tau)=e^{-iH_X\tau}e^{-iH_Y\tau}e^{-iH_Z\tau}=e^{-i H_F[\tau]\tau},
\end{equation}
where $\tau$ is the trotter step. If the frequency of the drive is high enough ($\tau$ is small), this evolution can be captured by an effective Hamiltonian $H_F[\tau]$ is approximated by the Magnus expansion~\cite{Magnus.1954}. For Eq.~\eqref{eq:Floquet}, the first two terms are
\begin{align}
    H_{F}[\tau] &= \underbrace{H_X+H_Y+H_Z}_{H_0} - \frac{i}{2}\tau[H_X,H_Y]\nn\\
    &-\frac{i}{2}\tau[H_X,H_Z]-\frac{i}{2}\tau[H_Y,H_Z] + O(\tau^2)\label{eq:Hf}.
\end{align}
This kind of Floquet Hamiltonian engineering by time averaging has been used in the past to construct prethermal phases including time crystals~\cite{Else.2017,Machado.2020,Yao.2017,Choi.2017}, and lattice gauge theories~\cite{Zohar.2017,Schweizer.2019}.
In these approaches, the terms other than $H_0$ are typically detrimental to the desired evolution and must be suppressed. In contrast, we find that the first-order terms $O(\tau)$, explicitly written in Eq.~\eqref{eq:Hf}, are crucial for an efficient realization of the desired non-abelian phase, as discussed below.

The leading-order term,
\begin{equation}\label{eq:H0}
    H_0=H_X+H_Y+H_Z,
\end{equation}
realizes the \emph{Kitaev honeycomb} model~\cite{Kitaev.2006}, which describes a highly anisotropic spin system on a honeycomb lattice [Fig.~\ref{fig:1}(a)]. This model features several gapped abelian topological phases when $\abs{J_{\alpha}} > \abs{J_{\beta}}, \abs{J_{\gamma}}$, and a gapless non-abelian phase where the linear dispersion is protected by the time-reversal symmetry. That symmetry needs to be broken in order to introduce a gap in the spectrum, which is necessary for a finite correlation length of the excitations thus making them localizeable.
The first-order dynamical correction to the effective Hamiltonian,
\begin{equation}
    H_1 = -\frac{i\tau}{2}([H_X,H_Y]
    +[H_X,H_Z]+[H_Y,H_Z]),
\end{equation}
which consists of three-body operators of the form $X_iY_jZ_k$, breaks the time-reversal symmetry since the signs depend on the particular ordering of applied Hamiltonians in~\eqref{eq:Floquet}; here, $H_X,H_Y,H_Z$. 
Because the gapped phase disappears in the static, high frequency limit $\tau \rightarrow 0$, 
the effective Hamiltonian realizes an intrinsically dynamical \emph{Floquet spin liquid}.
In the latter part of this work, we develop more sophisticated sequences to further suppress higher-order terms in Eq.~\eqref{eq:Hf}, enabling coherent control of the particles, while at the same time, controlling the relative signs and magnitudes of the desired three-body terms $H_1$. We discuss this procedure, and a more specialized approach for the state preparation, in Sec.~\ref{sec:variational}. Similar dynamical constructions have been used to propose Floquet SPT phases~\cite{Potirniche.2016} and the realization of chiral edge phenomena~\cite{Po.2017}.

The Kitaev Hamiltonian $H_0+H_1$ has an extensive number of conserved quantities given by plaquette operators [Fig.~\ref{fig:1}(b)],
\begin{equation}\label{eq:W}
W_p = Y_1 Z_2 X_3 Y_4 Z_5 X_6,
\end{equation}
for each plaquette $p\in\{1...M\}$ where $M=N/2$ is the total number of plaquettes and $N$ is the number of spins in the system. The collection of values $\mathcal{W}=\{W_p=\pm 1 \,|\, p\in 1...M\}$ defines a symmetry sector and it can be shown that the ground state lies in the sector where $W_{p}$\,$=$\,$+1$ for all $p$~\cite{Lieb.1994}. We associate the presence of a $\mathbb{Z}_2$ vortex on the plaquette $p$ if $W_p$\,$=$\,$-1$ and call the collection $\mathcal{W}$ a ``vortex sector''; in this language the ground state is vortex-free. Even though the number of symmetries is extensive, the vortex configuration does not specify the wavefunction exactly and each sector has a Hilbert space of dimension $2^{N/2}$.

Before proceeding we note that in many previous proposals, including the original work of Kitaev~\cite{Kitaev.2006}, the time-reversal symmetry was broken in a perturbative fashion---either by applying an external magnetic field~\cite{Kitaev.2006} or introducing ancillary gadgets to mediate interactions~\cite{Taylor.2011}. 
A key advantage of our dynamical scheme is that the time-reversal symmetry is broken without creating vortices: Each driving term in~\eqref{eq:Floquet} separately commutes with the plaquette operators, $[H_{Z,X,Y},W_p]=0$, so the effective Hamiltonian commutes as well $[H_F[\tau],W_p]=0$, irrespective of $\tau$. 
Additionally, each of the Hamiltonians $H_{X,Y,Z}$ can be written as a free fermion Hamiltonian~\cite{Kitaev.2006}, which means the heating due to trotterized time evolution is significantly suppressed and does not grow indefinitely~\cite{Lazarides,Gritsev}; see Appendix~\ref{app:heating} for more details.
This makes our Floquet spin liquid especially attractive for experimental implementation, where trotter heating can be a limiting factor.
The implementation can also be made hardware efficient, especially on reconfigurable Rydberg atom arrays, where dynamically toggling between $H_{X,Y,Z}$ is a natural operation~\cite{Bluvstein.2022}.

\section{Experimental Implementation}

Recently,  Rydberg atom arrays have been used to implement analog simulation of quantum spin systems with Ising~\cite{Ebadi.2021,Scholl.2021} and XXZ~\cite{Scholl.2022} Hamiltonians, using programmable,   
fixed atom geometry.
Most recently, a new architecture introduced coherent and parallel transport of atoms allowing for dynamically re-configurable connectivity~\cite{Bluvstein.2022}. 
In this approach, qubits are encoded in two magnetically insensitive hyperfine (HF) states $\ket{0}$,$\ket{1}$ while excitation to the Rydberg state $\ket{r}$ is used for performing entangling gates~[Fig.~\ref{fig:1}(e)]. This encoding provides long coherence times, enabling the execution of tens of thousands of parallel moves. Even for hundreds of qubits,  the quantum gates can be performed by illuminating the atoms with a global laser beam which provides a native parallelism of quantum operations with no cross-talk.  Therefore, reconfigurable Rydberg atom arrays are easily scalable and especially well suited for performing parallel quantum circuits where during each step, or \emph{gate layer}, each qubit participates in at most one gate.

Here we describe how the Floquet unitary~\eqref{eq:Floquet} can be implemented in this experimental platform in a hardware-efficient manner.
Since the honeycomb lattice is bipartite, every link connects the odd (\fcircle) and even (\ecircle) sublattices; see Fig.~\ref{fig:1}(a). Therefore, the evolution under each of the two-body Hamiltonians $H_X$,$H_Y$,$H_Z$ can be implemented within a single gate layer using a global pulse that simultaneously acts on all qubits. This pulse performs the two-body gate $\mathcal{G}_2(\theta)=e^{i\theta Z Z}$, which is equivalent to a controlled phase gate up to local $Z$ rotations. 
First, the atoms are all transported in parallel and brought together along the $XX$ links [see Fig.~\ref{fig:1}(e)], and a Raman pulse performs a global single-qubit gate that changes the basis from $Z$ to $X$ in the HF manifold. Then, the entangling gate operation $\mathcal{G}_2$ is applied to all pairs in parallel, which effectively performs $e^{-i H_X \tau}$. 
The atoms are then transported again to the next link configuration and the procedure is repeated, with an appropriate basis change in-between, requiring at all times only a single family of entangling gates (controlled phase gates). The dynamical change of connectivity in a system of 10s-100s of qubits available in Rydberg atom arrays is central to the realization of our Floquet protocol. Moreover, such programmability allows us to implement periodic boundary conditions without significant overhead, which removes gappless edge modes that obstruct ground state preparation under open boundary conditions.

The characteristic coherence time of such a digital simulation depends on the error rates of the constituent components.
In particular, since the global single-qubit gates within the HF manifold and the  transport of atoms have negligible effect on the coherence of stored qubits~\cite{Sheng,Levineb8q,Bluvstein.2022}, the errors will be dominated by the two-qubit operations. Most recently, control-Z Rydberg gates with fidelities beyond $99.5\%$ have been demonstrated~\cite{Evered.2023}, and are expected to go above $99.9\%$ in the near future. Moreover, the control-Z gate is a special case of the $\mathcal{G}_2$ gate ($\theta$\,$=$\,$\pi/4$) and similar gates with smaller angles are faster and therefore expected to work at even higher fidelities.

While the implementation in reconfigurable Rydberg tweezer arrays is particularly efficient, the scheme described in the this work can be realized in other quantum simulation platforms.  For example, trapped-ion quantum charge-coupled devices~\cite{pino2021demonstration,Iqbal} support dynamical connectivity while superconducting-qubit devices allow for selective application of two-body gates~\cite{Google.2021}, both of which can be used to realize the Floquet cycle as described here.
\begin{figure*}
    \centering
    \includegraphics[width=\linewidth]{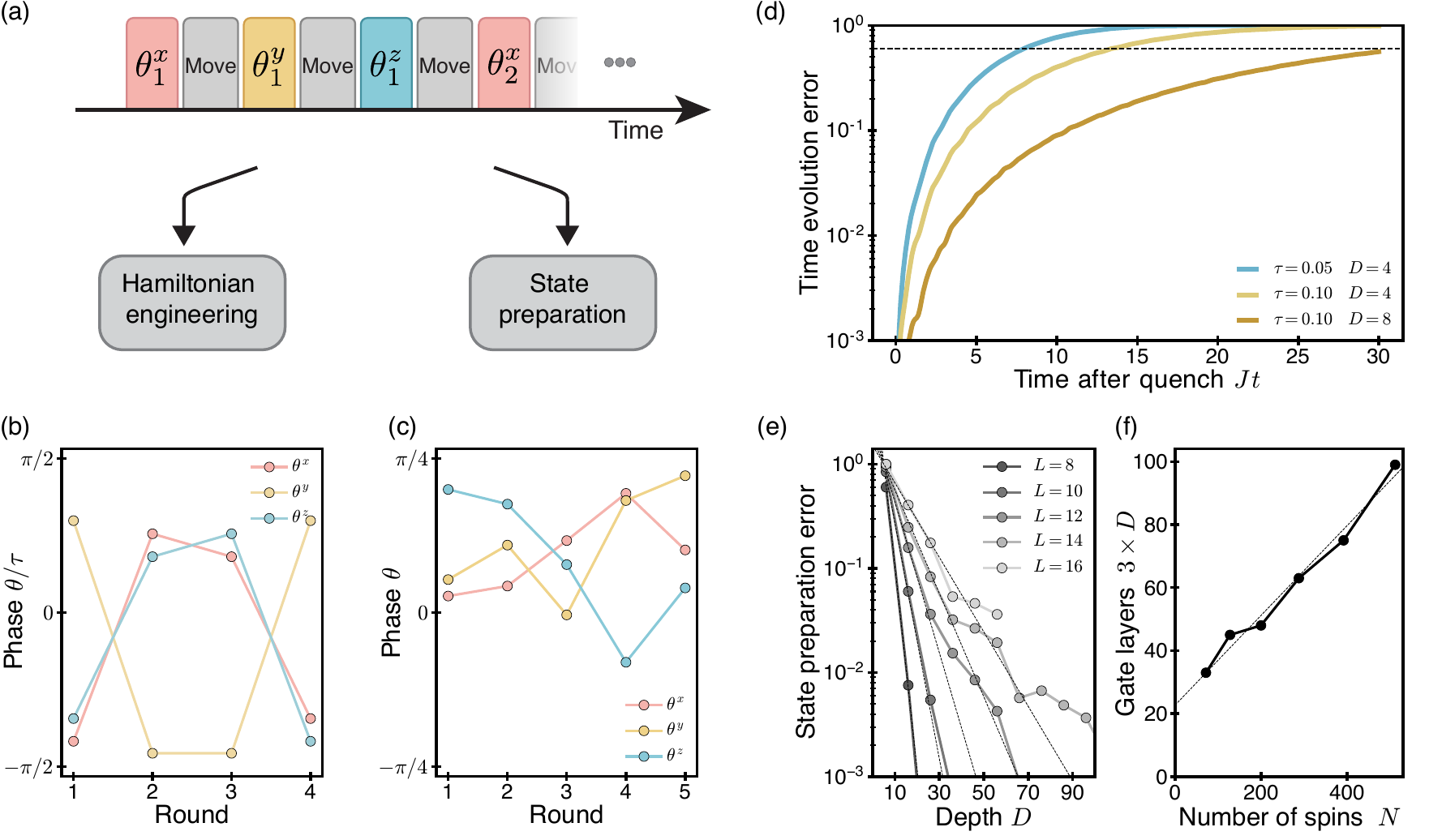}
    \caption{\textbf{Variational circuits for Hamiltonian engineering and state preparation.} (a) The circuit ansatz from Eq.~\eqref{eq:circ} suffices for efficient approximation of the target Hamiltonian as well as finding specialized circuits for state preparation. The variational circuit yields pulse sequences that result in Floquet Hamiltonians where both the strength and signs of couplings can be controlled. Preparing a certain output state is less demanding then reconstructing a many-body operator; thus, with the same variational ansatz, we can prepare the non-abelian ground state more efficiently than performing adiabatic evolution both in terms of the gate count and resulting fidelities. (b-c) Phases obtained through variational optimization on a relatively small 30-qubit system realizing effective Hamiltonian evolution and variational state preparation, respectively.
    (d) For engineered Floquet Hamiltonians, longer pulse sequences can systematically improve the approximation to the target Hamiltonian $H$. (e) Performance of the variational state preparation on an $L\times L$ lattice. Error in the final state as a function of the circuit depth $D$ for several system sizes. The error decreases exponentially with $D$ as $1$\,$-$\,$\mathcal{F}$\,$\propto$\,$\exp(-AL^{-\alpha}(D-D_0))$ for low depths; we estimate $\alpha$\,$\approx$\,$2.43$, $A$\,$\approx$\,73, and $D_0$\,$=$\,5. (f) The number of two-qubit gate layers necessary to achieve the many-body state fidelity of $90\%$ for a system of $N=2L^2$ spins.}
    \label{fig:2}
\end{figure*}

\section{Variational Optimization for Near-term Devices}\label{sec:variational}
In this section, we describe a procedure to systematically construct high-fidelity pulse sequences for resource efficient Hamiltonian simulation and state preparation through variational optimzation, guided by physical insights and experimental simplicity.

As described in Eq.~\eqref{eq:Hf}, the symmetry-breaking three-body terms appear in the first order of the Magnus expansion in the Floquet unitary~\eqref{eq:Floquet}. Thus, the repeated application of this unitary, with an appropriate value of $\tau$, should already resemble the time evolution under $H$ to a certain degree. We can further improve the fidelity if we promote the individual phases to variational variables which we then optimize. Concretely, the circuit ansatz of depth $D$ consists of $D$ blocks based on the Floquet unitary in Eq.~\eqref{eq:Floquet},
\begin{equation}\label{eq:circ}
 U_D(\{\theta\}) = \prod_{i=1}^D U_b(\vec{\theta}_i),
\end{equation}
where $\{\theta\}$ enumerates the set of all variational parameters. The $i$th block, $U_b(\vec{\theta}_i)$, is
\begin{equation}
    U_b(\vec{\theta}_i) = e^{i\theta^x_i XX}e^{i\theta^y_i YY}e^{i\theta^z_i ZZ},
\end{equation}
where $XX$, $YY$, $ZZ$ denote the respective two-qubit gates applied to all appropriate links in parallel---this corresponds to Eq.~\eqref{eq:Floquet} with the couplings promoted to variational variables. Thus, a circuit ansatz of depth $D$ consists of $3D$ two-qubit gate layers interleaved with global single-qubit rotations.

Within a fixed vortex sector, the Floquet dynamics are exactly solvable and the individual terms correspond to quadratic, free-fermion operators; see Ref.~\cite{Kitaev.2006} and Appendix~\ref{app:maj}. This enables efficient analytical/numerical treatment of the circuit in Eq.~\eqref{eq:circ} including calculation of gradients $\partial_{\vec{\theta}}U_D$. In particular, we can simulate a system of $N$ sites by working with matrices of size $N \times N$.
Then, we are free to optimize the phases $\{\theta\}$, and hence the pulse sequence, using our choice of a numerical optimization method. The cost function $Q$ to be minimized would depend on the specific application. Moreover, the translational invariance of the system enables us to analytically evaluate low-order terms in the Magnus expansion. In the next section, we use it to build intuition underlying good solutions to the variational problem.

\subsection{Effective Hamiltonian}
First, we focus on constructing effective Floquet Hamiltonians using the variational ansatz above. 
Specifically, we minimize the cost function
\begin{equation*}
	Q_{H} = \norm{U_D(\{\theta\})-e^{-iH\tau}}_F^2
\end{equation*} 
using gradient descent starting from a randomized symmetric configuration of phases, as described below. 
The norm we use is the squared Frobenius norm $\norm{A}_F^2=\tr{AA^\dagger}$, which is amenable to simple gradient calculation; see Appendix~\ref{app:Heff} for details. 

An example pulse for $\tau=0.05$ and $D=4$ is presented in Fig.~\ref{fig:2}(b).  Here, the general structure can be understood as the optimizer fixing the correct signs and magnitudes of the target three-body terms. The intuition that such a symmetric configuration of phases is enough for this purpose can be inferred from a simpler $D=2$ case where we parameterize all phases with two variables $\phi,\delta$,
$U_2=e^{i(\phi-\delta)H_X}e^{i\phi H_Y}e^{i(\phi+\delta)H_Z}e^{i(\phi+\delta)H_X}e^{i\phi H_Y}e^{i(\phi-\delta)H_Z}$. The first two orders of the Magnus expansion are $H_F^{(0)} = 2\phi H_0$ and
\begin{align*}
    H_F^{(1)} &= \phi(\phi-\delta)[H_X,H_Y]\\
    &+ (\phi^2-2\phi\,\delta-\delta^2)[H_X,H_Z]\\
    &+ \phi(\phi-\delta)[H_Y,H_Z].
\end{align*}
In particular, by choosing $\delta$ appropriately, we can make the $[H_X,H_Z]$ coefficient negative while keeping $[H_Y,H_Z]$ and $[H_X,H_Y]$ positive. Demanding the two have opposite signs and equal magnitude---which guarantees uniform couplings when combined with the signs from commutators---we get the condition 
\begin{align*}
    0 &= 2\phi^2 - 3\phi\delta -\delta^2& \Rightarrow& &
    \delta &= (-3\pm\sqrt{17})\phi/2,
\end{align*}
which we can straightforwardly solve. This simple example shows that we can freely tune the strength and signs of the three-body terms arising in $H_F^{(1)}$. 

Unfortunately, reducing the contribution of higher-order terms $H_F^{(n>1)}$ while preserving the first two is a formidable challenge and we  empirically find that the variational landscape has many local minima and plateaus making the optimization result strongly dependent on a good initial seed. However, it is much easier to find good $H_F[\tau]$  for small $\tau$ and thus the following recursive procedure turns out to  be  successful: After obtaining a good approximation to $H_F[\tau]$ at depth $D$, we apply the $e^{-iH_F[\tau]\tau}$ unitary twice and use it as an initial guess for a depth-$2D$ approximation to $H_F[2\tau]$.  This corresponds to evolving the system for time $2\tau$ and asking the optimizer to correct the combined discrepancies, which intuitively should be perturbatively small in $\tau$ and thus amenable to gradient-based methods.

The results are presented in Fig.~\ref{fig:2}(d) where the many-body evolution is simulated with various effective Hamiltonians and the failure rate of this process is measured via the many-body state overlap as $1-\abs{\braket{e^{iH_F[\tau]t}e^{-iHt}}}^2$. As expected, the $H_F[\tau$\,$=$\,$0.1]$ evolution (yellow) at $D$\,$=$\,$4$ outperforms the $D$\,$=$\,$4$ simulation with $H_F[\tau$\,$=$\,$0.05]$ (blue) by a factor of 2 in the time it takes to reach a certain threshold (dashed line); intuitively, this is consistent with the $\tau$\,$=$\,$0.05$ circuit requiring twice the number of gates for the same physical time $J\,t$. However, the $D$\,$=$\,$8$ Hamiltonian $H_F[\tau$\,$=$\,$0.1]$ (brown) obtained by the above-mentioned recursive construction allows for 6 times longer time evolution while requiring the same number of total applied gates  as the $\tau$\,$=$\,$0.05$ case. While the improvement is not surprising,  we found that already at circuit depth $D$\,$=$\,$8$ it is difficult to find reasonable solutions without guided initial state and thus this iterative procedure is essential for finding good Hamiltonians at larger $\tau$, allowing for long evolution time. The general task of finding optimal pulse sequences is an interesting optimization problem and could be a fertile ground for quantum signal processing~\cite{10.1103/prxquantum.2.040203} and machine learning methods, which have been used in the past to successfully engineer robust pulse sequences for nuclear magnetic resonance (NMR)~\cite{Peng.2021}.

\subsection{Variational state preparation}
In principle, an efficient Floquet Hamiltonian enables both the energetically protected operations on the non-abelian excitations and adiabatic state preparation. However, each Hamiltonian evolution step requires several gate layers, which limits the total available evolution time. We find that the state preparation step can be significantly improved, compared to the adiabatic state preparation, by employing the same variational methods used to develop efficient Floquet Hamiltonians. Intuitively, it is easier to transform a given initial vector to the final one (state preparation) compared to approximating many-body evolution on the full Hilbert space (effective Hamiltonian construction). 
Again, we use the same circuit ansatz from Eq.~\eqref{eq:circ} but instead of minimizing a Frobenius norm of the Hamiltonian evolution we choose to use the many-body state overlap as the cost function for the optimizer,
\begin{equation*}
	Q_{\rm VSP} = -\abs{\braket{\psi_{\rm fin}|U_D(\{\theta\})|\psi_{\rm ini}}}^2,
\end{equation*}
where $\psi_{\rm ini}$ and $\psi_{\rm fin}$ are the initial and target states, respectively. This cost function and its gradient can also be evaluated efficiently using the free-fermion picture; see Appendix~\ref{app:Heff} for details.

In Fig.~\ref{fig:2}(e), we present the performance of variational state preparation (VSP) for $\psi_{\rm ini}$ and $\psi_{\rm fin}$ being the vortex-free ground states of $H_Z$ and $H$, respectively. The ground state of $H_Z$ is obtained by projecting a product state $\psi_0$ into the no-vortex symmetry sector, which can be done in a single step by performing projective measurements of $W_p$ operators or, alternatively, by passively cooling the state using conditional gate operations which progressively remove vortices; we describe these procedures in Appendix~\ref{app:proj}. The task accomplished by the VSP corresponds to the transition from the toric code (the abelian phase) to the non-abelian chiral phase. We find that the error in the preparation of the target state decreases as $\exp(-DL^{-\alpha})$ in the low-depth regime, where we numerically estimate $\alpha\approx 2.43$ and $L=\sqrt{N}$. Remarkably, even though the circuit depth scales with the system size---as required since crossing a phase transition involves building up long-range order---the depth necessary for state preparation is accessible in Rydberg atom arrays even for hundreds of spins. We note that a similar state preparation scheme, although not employing the Floquet stabilization of the non-abelian phase nor experimental co-design, has been explored in Ref.~\cite{Bespalova.2021}.

The many-body fidelity is convenient for optimizing the performance of the VSP but it is not very useful for experimentally assessing the quality of the final state and its robustness to errors. For this purpose, we additionally benchmark the state preparation under a control noise model. We simulate a VSP circuit, which prepares a logical state encoded in the ground state, with the optimal phases modified by errors sampled from a uniform distribution. We then read out the logical state with our dynamical quench protocol (discussed later) and the outcome allows us to quantify the quality of the prepared state in an experimentally feasible way. We find that even imperfect state preparation results in a high probability of correct measurement outcome, up to large phase fluctuations of $\delta\theta$\,$\approx$\,$0.06$. This suggests that our state preparation procedure is, to a certain extent, robust against these types of errors; the details of this benchmark can be found in Appendix~\ref{app:noisyvsp}.

We also point out a subtlety related to the different ground-state degeneracy of the abelian and non-abelian phases: The dimension of the ground-state manifold in the toric code phase is 4, while for the non-abelian phase it is 3. This means that one of the ground states is not adiabatically connected to the non-abelian theory and becomes an excited state. By choosing an appropriate initial product state $\ket{\psi_0}=\ket{0}^N$, we ensure that the system is orthogonal to this undesired state. We discuss this important point in more detail in Appendix~\ref{app:prep}.

Finally, we note that Hamiltonian learning methods~\cite{Kokail.2021,Pastori.2022} could be used to confirm the preparation of the non-abelian phase without the need for sophisticated control of the excitations. Similarly, variational state preparation combined with effective Hamiltonian evolution suffice to probe the chiral nature of the edge modes. However, performing basic TQC operations, such as braiding and fusion measurements, not only verifies the non-abelian nature of the particles but also paves the way to more complex computing tasks. In the rest of this work we focus on describing steps necessary to achieve these goals.

\section{Controlling non-abelian excitations}
\begin{figure}
    \centering
    \includegraphics{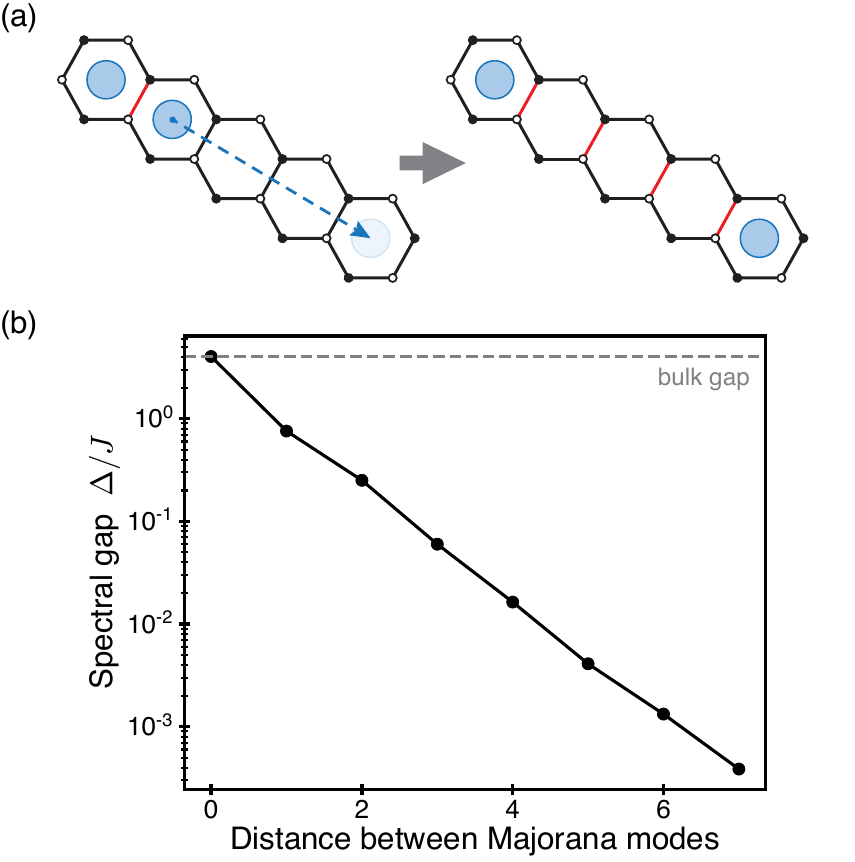}
    \caption{\textbf{Majorana zero modes.} Movement of non-abelian modes can apply non-trivial unitary operations to the logical subspace. (a) Two modes localized on adjacent plaquettes are created when the sign of a single bond coupling is adiabatically flipped (red). The flipping of subsequent bonds in a ladder-like pattern allows for their transport. (b) As the modes are moved away from each other, their energy decreases exponentially fast with the separation distance, eventually resulting in the Majorana zero modes. The degenerate \emph{logical subspace} encodes the protected qubit: The empty mode corresponds to $\ket{0}_L$ and the occupied one to $\ket{1}_L$. See also Fig.~\ref{fig:1}(d).}
    \label{fig:3}
\end{figure}
One of the remarkable properties of the Kitaev honeycomb model is the presence of excitations that exhibit non-abelian statistics, specifically of the Ising anyon type~\cite{Nayak.2008}.
Unlike the abelian anyons present in the toric code model~\cite{Kitaev.2003}, operations on the non-abelian $\sigma$ particles result in unitary matrices---not only a global phase---acting on the logical qubit subspace. 
In this section, we present the steps necessary to create, manipulate, and read out the non-abelian $\sigma$ particles, which hold Majorana zero modes. 
For the rest of this work, we will use the terms ``Majorana zero mode'' and ``$\sigma$ particle'' interchangeably.

\subsection{Creating and moving zero modes}
To create Majorana zero modes, first the sign of a single bond can be flipped adiabatically, which affects the $J$ coupling and the overlapping three-body terms in the extended Hamiltonian $H$, without changing the vortex sector $\mathcal{W}$~\cite{Taylor.2011}.
This modifies the spectrum, turning one of the excited fermion eigenstates states into a low-energy bound state localized around the flipped bond.
As a chain of adjacent bonds are adiabatically flipped, the energy of the original fermion state decreases exponentially with distance, rapidly approaching a zero energy mode that can be occupied at no cost. This forms a nearly-degenerate two-dimensional ground space. 
Furthermore, local measurements can no longer determine whether the originally excited state is occupied or unoccupied; hence these states form topologically protected ``logical states'' ($\ket{0}_L$ and $\ket{1}_L$) which we identify with sectors $1$ and $\psi$ respectively.

Taken together, this degenerate subspace can also be interpreted as two Majorana zero modes, supported on plaquettes connected by the flipped bond [Fig.~\ref{fig:3}]. 
Since the global fermion parity is conserved, the simplest non-trival system consists of four Majorana modes, which span a two-dimensional logical subspace.
For a given pairing, these two states correspond to no fermions ($11$) or two fermions present ($\psi\psi$) [Fig.~\ref{fig:1}(d)]. 
This two-level system can be manipulated by transporting Majorana modes. This can be done in a topologically and energetically protected way by adiabatically flipping the signs of appropriate bonds using the effective Hamiltonian evolution.

Our Floquet approach to stabilizing the excitations has the crucial advantage that a local modification of phases in the variational ansatz~\eqref{eq:circ} automatically modifies, to leading order, the emergent three-body terms in an appropriate manner. The variational circuit is straightforwardly extended to include such local modifications, and can be optimized at every step if necessary.
Experimentally, local phases can be realized either with local Rydberg laser control or by separating local terms into separate gate layers.

\subsection{Localized fermion readout}\label{sec:readout}
\begin{figure}[tb]
    \centering
    \includegraphics{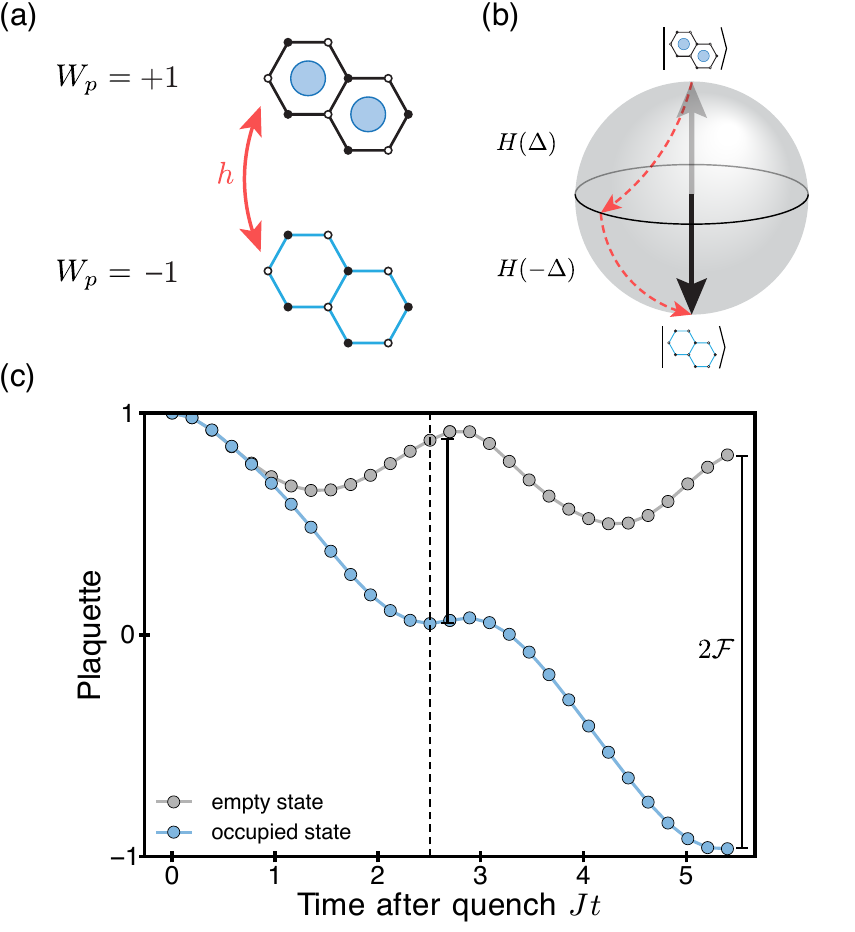}
    \caption{\textbf{Readout of the localized fermion.} (a) Magnetic field $h$ creates/annihilates the localized fermion mode and flips the neighboring plaquettes. This allows for mapping the presence of a local excitation to a spin observable. (b) The effective two-level system is conveniently represented on a Bloch sphere. We use a two-part pulse consisting of the Rabi drive under the Hamiltonian $H(\Delta)$ and the second equally long evolution under $H(-\Delta)$ with an opposite sign of the detuning. (c) The plaquette expectation value for the two states of interest under evolution with the two-stage pulse sequence.  The opposite detuning is achieved by utilizing the digital nature of the simulation and applying $-H$ instead of $H$. The use of the composite pulse increases the contrast from $\mathcal{F}=0.43$ (first pulse, dashed line) to $\mathcal{F}=0.89$ (both pulses).}
    \label{fig:4}
\end{figure}
The key step necessary to complete the particle fusion is the readout of the localized fermion state (either $1$ or $\psi$). In our case, these fermions are formed from two adjacent Majoranas, $\psi = (c_1+ic_2)/2$, brought together to perform fusion, as in the final step of Fig.~\ref{fig:5}(a).
In other words, we want to check whether the two Majorana modes are paired or not. In our figures we indicate such pairing checks with solid blue arrows.
The key insight is that the magnetic field in the $Z$ direction, applied to the bond shared by the adjacent Majorana particles, couples the $\psi$ particle to an empty state with two vortices [Fig.~\ref{fig:4}(a)]; i.e, it destroys a fermion and flips adjacent plaquettes. This state, without a localized fermion, is the ground state in the two-vortex sector.
Because the initial state couples strongly to another eigenstate, the two effectively form a detuned two-level system (TLS) with the detuning $\Delta$ on the order of the bulk gap; see Appendix~\ref{app:tls} for the details of this TLS model.
Alternatively, if the fermionic mode is initially empty, we create a localized fermion mode in addition to the two vortices. However, in the two-vortex sector, there is no localized-fermion state and the momentum eigenstates form a band with quadratic dispersion.
Hence, the empty state is weakly coupled to a detuned continuum, and the response to the local magnetic field quench is different than for the occupied state discussed above.

This discrepancy in the response allows us to differentiate between the two states in an experiment: We utilize the strong coupling to map the $\psi$ particle to the two $\mathbb{Z}_2$ vortices whose presence is read-out by measuring the adjacent plaquette operator $W_p$---a simple spin observable. This realizes a measurement within the TQC framework. It is also similar in spirit to phase measurements in AMO systems. We emphasize that our protocol does not depend on any fine-tuned parameters and can be calibrated independently in a small-scale experiment.

In Fig.~\ref{fig:4}(c), we show the time evolution of the relevant plaquette expectation value $\braket{W_p}$ after the quench. As expected, we observe a much stronger response during the quench from the occupied fermionic mode compared to the empty one. The figure of merit for the readout is the difference between the two cases 
\begin{equation}
    \mathcal{F} = \abs{\braket{W_p}_{\rm emp.}-\braket{W_p}_{\rm occ.}}/2,
\end{equation}
and, ideally, we could map the empty state to $\braket{W_p}_{\rm emp.}=+1$ and the occupied state to $\braket{W_p}_{\rm occ.}=-1$. That would allow us to distinguish them with perfect fidelity $\mathcal{F}=1$.
Unfortunately, the system undergoes detuned oscillations in the TLS, never achieving the maximal contrast. However, we can leverage the two-level analogy, and the flexibility of digital simulation, to devise robust composite drives that increase the fidelity dramatically. Such pulses have long been used in the NMR community. We find that a simple two-stage composite drive already increases $\mathcal{F}$ substantially. It consists of the time evolution until the first local minimum and the subsequent evolution for the same time albeit with a negative detuning [Fig.~\ref{fig:4}(b)]. We achieve the negative detuning by keeping the local magnetic field fixed while driving with $-H$, an easy task in our digital approach. In Fig.~\ref{fig:4}(c), the first pulse lasts until the dashed line and the use of the composite pulse improves the fidelity from $\mathcal{F}=0.43$ to $\mathcal{F}=0.89$.

Including effective magnetic fields in the model prevents the exact solution and makes the analysis of the system much more difficult. However, we can obtain an approximate solution using the fermionic Gaussian state (FGS) ansatz~\cite{Kraus.2010,Shi.2018}---this choice is well motivated here since eigenstates in well-defined vortex sectors are exactly Gaussian. Hence, we expect this variational ansatz to interpolate well between them. We find that the FGS predictions are in good agreement with the exact results for small system sizes and short to intermediate times; see Appendices~\ref{app:fgs} and~\ref{app:rcal} for details.

In the effective Hamiltonian, we include the magnetic field  by performing a single-qubit $Z$-phase gate $e^{i\theta Z}$ at the appropriate site; since the field is local, this first-order treatment is enough to approximate the full Hamiltonian evolution. Finally, because the readout procedure depends only on local quenches, it can be calibrated in a relatively small 30-qubit experiment where the degree of control over the state is much larger, and subsequently used in the full-scale simulation. We describe such a protocol in detail in Appendix~\ref{app:rcal} and supplement it with exact spin-picture simulations. We note that going beyond quench dynamics, and optimizing the performance of this readout scheme with a variational procedure, is an interesting direction for future work.

With these tools, allowing for the preparation, movement, and readout of the excitations, it should now be possible to perform the two basic experiments characterizing an anyonic theory: Fusion and braiding. We describe them briefly in the following section.

\section{Fusion and Braiding}
\begin{figure}
    \centering
    \includegraphics{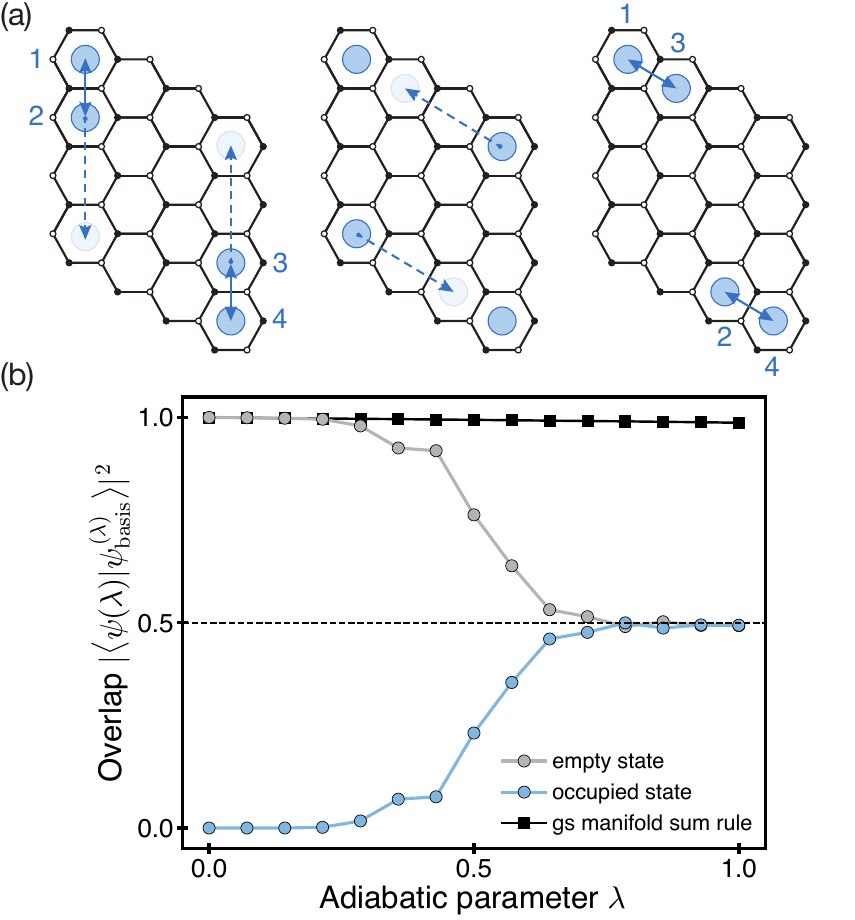}
    \caption{\textbf{Fusion of non-abelian anyons.} (a) Two pairs of Majorana modes (1,2 and 3,4) are created in the logical $\ket{0}_L$ state (empty zero mode) and subsequently moved and brought together in a different configuration. During each step the modes are moved a single site by adiabatically flipping the sign of the subsequent bond. (b) Many-body overlap with instantaneous logical basis states along the adiabatic path. The $\lambda$\,$=$0, 0.5, and 1.0 correspond to initial, intermediate, and final state of the fusion experiment depicted in (a). This adiabatic process corresponds to a change of logical basis or, equivalently, applying a Hadamard gate. The simulation was performed on an $L\times L$ lattice with $L$\,$=$20 with an idealized, target Hamiltonian. In practice, this will be realized using Floquet circuits (see Fig.~\ref{fig:2}). The modes are initially placed on neighboring plaquettes and the two pairs are separated by $L/2-2$, which results in the total of $2(L/2-2-1)=14$ steps necessary to complete the fusion. For braiding at a characteristic distance $\lfloor\frac{L}{3}\rfloor$, the required number of steps is $9\lfloor\frac{L}{3}\rfloor-23$.}
    \label{fig:5}
\end{figure}
First, we consider a procedure to confirm the non-abelian fusion rule,
\begin{equation}\label{eq:fusion}
    \sigma\times\sigma\to1+\psi,
\end{equation}
corresponding to the recombination of two $\sigma$ particles. Unlike the toric code case, where combining two particles always gives a deterministic result, here fusing $\sigma$ particles can, under certain circumstances, produce either a vacuum state or a $\psi$ fermion, with equal probabilities. This behavior is a hallmark of a non-abelian theory~\cite{Nayak.2008}.

We initialize the logical $\ket{0}_L$ state by preparing two pairs of unoccupied Majorana modes (labeled 1-2 and 3-4) using the adiabatic protocol described in the previous part. Next, we move modes 2 and 3 in such a way that they recombine in the opposite pairing [Fig.~\ref{fig:5}(a)]; i.e., 1-3 and 2-4. In the language of TQC, this operation corresponds to the measurement in an orthogonal basis, e.g., $\{\ket{+}_L,\ket{-}_L\}$, and thus we expect both results to appear with equal probability. Physically, along the adiabatic path, the states become exponentially degenerate (zero modes) and thus the transition from one configuration to another is maximally diabatic, resulting in equal superposition of the final states $\ket{11}+\ket{\psi\psi}$. The evolution of the state can be traced by monitoring the state's decomposition in terms of the instantaneous logical basis [Fig.~\ref{fig:5}(b)]. Finally, we check for the presence of the $\psi$ fermion with our dynamical quench protocol. In experiment, measuring equal weights of the vacuum and $\psi$ states, combined with the high correlation between the two pairs, would confirm the non-abelian fusion rule~\eqref{eq:fusion}.

Fusion rules do not specify an anyonic theory completely. On top of it, we need to identify the braiding properties of excitations; i.e., the transformations of the degenerate state manifold upon moving anyons around each other. 
As we have discussed in the case of fusion, swapping the modes 2 and 3 results in the application of a $\pi/2$ rotation on a Bloch sphere, or equivalently a change of basis. Now, performing this operation twice (in a way that results in mode 2 looping around mode 3) corresponds to applying the $\pi/2$ rotation twice---resulting in a $\pi$-pulse that encodes the 
\begin{equation}\label{eq:ubraid}
    U_{\rm braid}=\begin{pmatrix}0&1\\1&0\end{pmatrix}
\end{equation}
operation in the logical manifold up to global phases. The initial configuration is the same as the fusion experiment, but now mode 2 is looped around mode 3 and the logical state is measured in the original basis; i.e., 1-2 and 3-4. At the end, we end up in the starting configuration, but a non-trivial braid has been completed between the Majorana particles.

After completing the braiding operation, we should end up in a flipped state: $\ket{0}_L\to\ket{1}_L$, which can be verified by performing the readout and measuring the state $\ket{1}_L$ with probability close to unity. This concludes the characterization of the non-abelian properties of the excitations in the system. 

\section{Resource estimates and experimental feasibility}
Next, we discuss resources necessary to implement our proposal in quantum hardware. In Rydberg atom arrays, two-qubit gate fidelities are the dominant source of error, compared to extremely efficient single-qubit gates and atom transport~\cite{Sheng,Levineb8q,Bluvstein.2022}, and thus we focus on the number of two-qubit gate layers necessary. We summarize our findings in Table.~\ref{tab:resource}, where we assume a time step $J\tau$\,$=$\,0.25 for the time evolution and the readout procedure; this value is large enough to enable efficient implementation while still ensuring good performance of fusion and braiding operations. We emphasize that even larger trotter angles do not necessarily lead to the breakdown of the topological phase due to the free-fermion nature of the honeycomb model, see Appendix~\ref{app:heating} for details.
\begin{table}[ht]
\caption{\label{tab:resource}\textbf{Gate-layer estimate for the proposal.} The number of required two-qubit gate layers varies between the different steps of the proposal. Several ingredients, such as state preparation, can be realized in current experimental platforms. The estimate is performed for a system size with $N$\,$=$\,$2L^2$ qubits and the time evolution implemented with depth-$D$ circuit. }

\begin{ruledtabular}
\begin{tabular}{l|c|c}
Stage & Number of gate layers & Source\\ \hline
Initial projection & 6 & App.~\ref{app:proj}\\
State preparation  & $22+0.3L^2$ & Fig.~\ref{fig:2}(f) \\
Time evolution (1 step)  & 3$D$ &Eq.~\eqref{eq:circ}\\ 
Topological readout  & 60$D$ & Fig.~\ref{fig:4}\\\hline
Total (edge dynamics) & 28 + 0.3$L^2$+30$D$  & App.~\ref{app:chiral}\\
Total (fusion) & 28 + 0.3$L^2$+12$D(L-1)$ & Fig.~\ref{fig:5}\\
Total (braiding) & 28 + 0.3$L^2$+12$D(9\lfloor \frac{L}{3}\rfloor-18)$ & Fig.~\ref{fig:5}\\
\end{tabular}
\end{ruledtabular}
\end{table}
The resource estimates for these circuits are quite favorable for near-term experimental exploration of the non-abelian phase.  
For example, preparation of the non-abelian phase in a 72-qubit system would consume only 33 two-qubit gate layers.
This remarkable efficiency is a direct result of the hardware-efficient encoding, utilizing simple Floquet sequences of parallel two-qubit gates and global single-qubit rotations to generate the higher-order terms that stabilize the non-abelian phase.

\begin{figure}[t]
    \centering
    \includegraphics{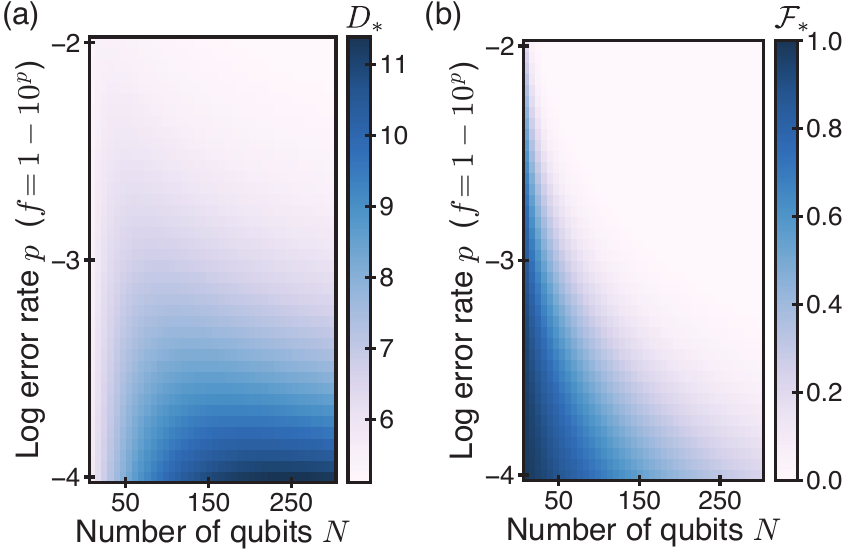}
    \caption{\textbf{Optimal circuit depth and fidelity in near-term devices.} (a) The optimal circuit depth $D_\ast$ maximizing the many-body fidelity during state preparation, balancing the trotterization and gate errors. The corresponding number of gate layers is $3D_\ast$. (b) The maximal many-body fidelity $\mathcal{F}_\ast$ at the given two-qubit gate fidelity $f$ and system size $N$, achieved at the optimal state-preparation circuit depth $D_\ast$. We note that the many-body fidelity is a stringent measure and many observables, such as topological readout, should be more robust to errors; see Appendix~\ref{app:noisyvsp} for further discussion.}
    \label{fig:6}
\end{figure}

Finally, we estimate the achievable many-body fidelity in near-term devices with faulty two-qubit gates. There are two competing processes: Longer sequences can ideally achieve higher state preparation fidelity, but accumulate more errors due to noisy entangling operations. To estimate the optimal circuit depth $D_\ast$, for a given two-qubit gate fidelity $f$, we use the heuristic scaling presented in Fig.~\ref{fig:2}(e) and assume an exponential decay of the many-body fidelity with the circuit volume,
\begin{equation}
    \mathcal{F}(D,f)\sim(1-e^{-A(D-D_0)L^{-\alpha}})f^{6L^2D}\label{eq:opt_fidelity},
\end{equation}
which allows us to calculate the optimal depth $D_\ast(f)$ and the corresponding maximal many-body overlap $\mathcal{F}_\ast$. 
In Fig.~\ref{fig:6}, we plot $D_\ast$ and $\mathcal{F}_\ast$ as a function of the system size $N$ and the gate fidelity $f$\,$=$\,$1$\,$-$\,$10^{p}$. Since current state-of-the-art devices can readily reach $f > 0.995$ and are projected to acheive $f > 0.999$ in the near-term, our estimate shows that significant many-body overlap with the topological ground state can be achieved in large systems of approximately a hundred qubits.
In practice, the many-body fidelity is the most  stringent measure and generically decays exponentially with system size. As such, signatures of the non-abelian topological phase should remain visible at larger system sizes as well.

\section{Discussion and Outlook}
The results presented in this work indicate that
non-abelian Floquet spin liquids can be dynamically created through applications of periodic pulse sequences in a hardware-efficient manner. 
In particular, we developed a dynamical protocol where the desired gapped Hamiltonian is obtained by engineering higher orders of the Magnus expansion and constructed efficient pulse sequences that approximate the target Hamiltonian and state preparation.
These tools can be utilized to encode logical information in the Majorana zero modes that can be adiabatically created and controlled with our effective Hamiltonian approach, providing 
a blueprint for exploring non-abelian phases of matter using Rydberg atoms arrays in the near-term. 
In particular, the Hamiltonian realization leads to energetic protection against coherent errors, such as control errors or spurious magnetic fields which induce local unitary rotations. In contrast, in a realistic simulation, incoherent errors will build up over time, unless removed by cooling. Developing hardware efficient cooling for the models studied here, for example by mid-circuit readout of stabilizers as in quantum error correction, may be important in practice for stabilizing the topological phase at long times.
The addition of such periodic readout intervals can also allow one to combine the present method with recently proposed Floquet quantum error-correcting codes, which dynamically realize topological phases via repeated measurements~\cite{Hastings_2021,Davydova_2022}.
Exploration of topological order in hybrid models with Hamiltonian evolution and mid-circuit measurements --- both of which can be realized using similar experimental controls --- is an intriguing future research direction.

The methods developed here can also be applied to study various extensions of the Kitaev model, which are the topic of current research, both theoretical and experimental. Adding external magnetic fields~\cite{Hickey.2019,Zhang.2022,Joy.2022}, Heisenberg interactions~\cite{Shinjo.2015,Joy.2022}, or more complicated many-body terms~\cite{Minami.2019}, may drastically change the behavior of the system. For example, different topological phases, with new classes of anyonic excitations, have been postulated for some of these extensions while others are especially relevant for real-world materials~\cite{Takagi.2019}. 
Additionally, we can leverage the efficient state preparation of the Kitaev honeycomb ground state and digital Hamiltonian evolution  to simulate quenches away from the integrable non-abelian phase to more complicated models, which are difficult to model on a classical computer. 
Such quenches enable the study of integrability breaking and thermalization near integrable points~\cite{Bertini.2015,DAlessio.2016,Znidaric.2020}.
Furthermore, it could also be used to probe excitations of the quenched Hamiltonian~\cite{Villa}, which is especially interesting due to the underlying topological order of the non-abelian phase.

Our approach can be extended to efficiently simulate other, more exotic models. For instance, an important class of lattice gauge theories can be efficiently simulated by implementing three-body interactions~\cite{Irmejs}, using   
a family of gates $\mathcal{G}_3(\theta)=\exp(i\theta Z_iZ_jZ_k)$ combined with sublattice rotations (see Appendix~\ref{app:lgt}). Such a family of three-qubit gates can be implemented  as a natural extension of $\mathcal{G}_2(\theta)$~\cite{  Evered.2023} since arranging atoms in a triangle guarantees a symmetric blockade, as discussed in Appendix~\ref{app:gates}. The speed and robustness to errors for such operations can likely be improved further using 
optimal control methods~\cite{Jandura.2022}. 
In the absence of an exact solution, several of the methods employed in this work would have to be modified. For example, the variational state preparation relies on the ability to calculate the many-body overlap in order to minimize the cost function. Instead, one would need to resort to the adiabatic state preparation or variational methods involving a quantum-classical feedback loop~\cite{Peruzzo.2014}. Simplifying these requirements and optimizing these methods is another  interesting direction for future research.

\begin{acknowledgements}
We thank D. Bluvstein, X. Gao, C. Kokail, R. Sahay, R. Samajdar, D. Sels, K. Xiang, and P. Zoller for useful discussions.
This work was supported by the US Department of Energy [DE-SC0021013 and DOE Quantum Systems Accelerator Center (contract no. 7568717)], the Defense Advanced Research Projects Agency (grant no. W911NF2010021), the National Science Foundation, and the Harvard-MIT Center for Ultracold Atoms.
N.M. acknowledges support from the Department of Energy Computational Science Graduate Fellowship under Award Number DE-SC0021110.
\end{acknowledgements}

\appendix

\section{Exact solution using Majorana operators}\label{app:maj}
In this Appendix, we summarize the solution of the extended Kitaev Hamiltonian using Majorana operators. We follow exactly the original derivation in Ref.~\cite{Kitaev.2006}.

Each spin degree of freedom is decomposed into four Majorana operators: $b^x,b^y,b^z,c$. Together they span a 4-dimensional Hilbert space $\widetilde{\mathcal{H}}$, which necessitates the projection to the physical 2-dimensional space $\mathcal{H}$. We achieve this by restricting ourselves to the $+1$ eigenspace of the projector
\begin{equation}
    D = b^xb^yb^zc,\label{eq:majproj}
\end{equation}
thus imposing $D=1$; i.e., $\ket{\psi}\in\mathcal{H}$ iff. $D\ket{\psi}=\ket{\psi}$. The physical spin operators (Pauli $\sigma^{x/y/z)}$) are represented with
\begin{equation}
    \sigma^{\alpha} = ib^\alpha c\label{eq:spmaj},
\end{equation}
which holds as long as they act on $\ket{\psi}\in\mathcal{H}$.

The Hamiltonian $H=H_0+H_1$ reduces in the Majorana picture to
\begin{equation}
    H = \frac{i}{4} \sum_{i\neq j = 1}^{N} A_{ij} c_i c_j,\label{eq:Hmaj}
\end{equation}
where $A_{ij} = 2 J_{ij}u_{ij}+2K_{ij}u_{ik}u_{kj}$ and $u_{ij}$ vanishes if the sites $i$,$j$ are not connected. In terms of the Majorana degrees of freedom, $u_{ij}=ib^{(\alpha)}_ib^{(\alpha)}_j=-u_{ji}$ and all of them commute with the Hamiltonian thus encoding the effective gauge degrees of freedom. We choose the ordering in such a way that $i$,$j$ belong to the odd (\fcircle) and even (\ecircle) sublattices, respectively. The plaquette operator $W_p$ from Eq.~\eqref{eq:W} is given by an oriented product of the $u_{ij}$ bonds around the plaquette.

The Hamiltonian~\eqref{eq:Hmaj} is quadratic in the Majorana operators, which means the quantum state is fully characterized by the skew-symmetric real matrix $\Gamma_{ij} = (i/2)\braket{[c_i,c_j]}$. The time evolution can be solved exactly and is given by
\begin{equation}\label{eq:majev}
    \Gamma(t) = U^\dagger(t)\Gamma(0) U(t),
\end{equation}
where $U(t) = \exp(-At)$. 

\section{Heating of the Floquet Kitaev spin liquid}\label{app:heating}
\begin{figure}
    \centering
    \includegraphics{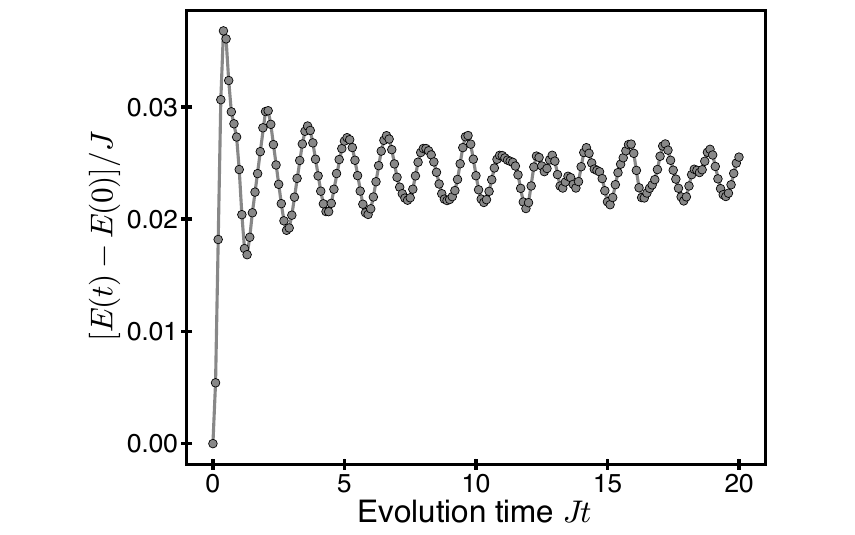}
    \caption{\textbf{Heating of the Floquet spin liquid.} Change in the expectation value of the target Hamiltonian $E(t) = \bra{\psi(t)}H\ket{\psi(t)}$ as a function of time. The energy rises sharply in short-time dynamics and eventually stabilizes to a value of the order $(J\tau)^2$. This behavior is a special feature of our free-fermion model and does not occur for generic interacting many-body systems.}
    \label{fig:sup_heat}
\end{figure}
Here we discuss heating  due to the intrinsic errors associated with trotterized Hamiltonian evolution. As shown in Appendix~\ref{app:maj}, the Kitaev honeycomb model is equivalent to a system of free fermions~\cite{Kitaev.2006}. Therefore, unlike generic many-body systems, it does not heat up indefinitely when subjected to external driving~\cite{Lazarides,Gritsev}. Instead, the energy of the system increases by a factor $\propto(J\tau)^2$ during short-time relaxation process and subsequently stabilizes without further heating, as presented in Fig.~\ref{fig:sup_heat}. 

Intuitively, the Floquet drive with frequency $\Omega = 2\pi/\tau$ introduces Floquet bands spaced at intervals of the order $\Omega$. If $\Omega$ is larger than the local energy scale, the heating becomes suppressed since correlated many-body processes are required to accommodate this large amount of energy. In other words, different Floquet bands need to overlap for the energy absorption to occur. In  generic interacting systems, the many-body bandwidth grows with the system size $N$ and, unless $\Omega \propto N$, different Floquet bands will eventually overlap leading to energy absorption and heating. In the free-particle system, however, the spectrum corresponds to the single-particle dispersion and is thus bounded regardless of the system size; this results in no heating as long as $\Omega$ is larger than that bandwidth. Since no energy quanta can be absorbed, the only effect of the drive comes from virtual processes which enter as $(J/\Omega)^2$ and dress local operators. 

Thanks to these special properties, the Kitaev honeycomb model is uniquely suited for the first large-scale demonstration of digital quantum simulation since the heating effects should be significantly suppressed compared to generic quantum systems. 

\section{Rydberg gates for Hamiltonian evolution}\label{app:gates}
Besides the application in this work, the $\mathcal{G}_2(\theta)=e^{i\theta ZZ}$ and $\mathcal{G}_3(\theta)=e^{i\theta ZZZ}$ families of gates allow for simulation of various exotic Hamiltonians such as lattice gauge theories. Here we introduce and characterize a simple realization of the $\mathcal{G}_3$ gate with three Rydberg pulses,
\begin{equation}\label{eq:G3}
    \mathcal{G}_3(\theta)=R(\Delta_1,\tau_1,2\phi)R(\Delta_2,2\tau_2,\phi)R(\Delta_1,\tau_1,0),
\end{equation}
where $R(\Delta,\tau,\phi)$ is a Rydberg pulse with the detuning $\Delta$, time $\tau$, and phase $\phi$, defined as the time evolution $e^{-iH\tau}$ under the Hamiltonian
\begin{equation}
    H = V_{\rm ryd}^{\infty}+\frac{\Omega\cos{\phi}}{2}\hat{X}-\frac{\Omega \sin{\phi}}{2}\hat{Y}  - \Delta\ket{r}\bra{r},
\end{equation}
where we use the energy units of $\Omega$, the time is in the units of $\Omega/2\pi$, and $V_{\rm ryd}^{\infty}$ denotes the perfect Rydberg blockade constraint; i.e., the $\ket{rr}$ and $\ket{rrr}$ states are infinitely detuned from the rest of the system. The two and three-body $W$ states arising due to the Rydberg blockade are defined as
\begin{align*}
    W_2 &= \frac{1}{\sqrt{2}}(\ket{1r}+\ket{r1}), \\ W_3 &= \frac{1}{\sqrt{3}}(\ket{11r}+\ket{1r1}+\ket{11r}).
\end{align*}
\begin{figure}
    \centering
    \includegraphics{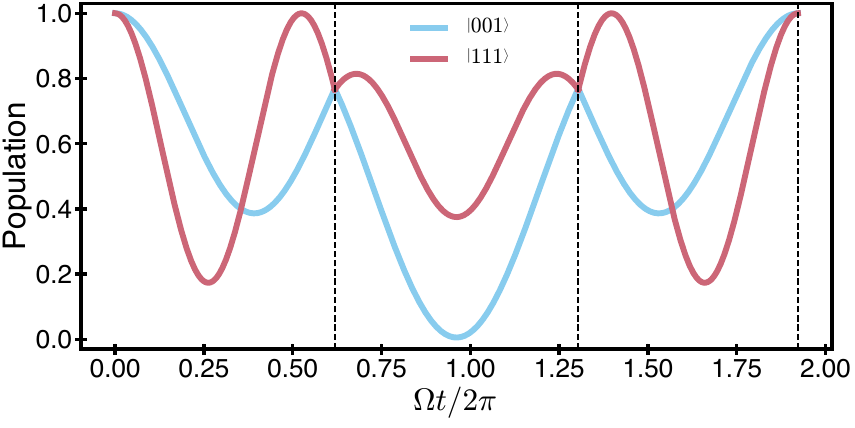}
    \caption{\textbf{Three-qubit gate: Populations.} Population dynamics during a pulse sequence realizing $\mathcal{G}_3(\theta=0.2)$. (a) Population of the hyperfine states during the gate execution. After every pulse (dashed line), the populations are equal between curves. Additionally, the second pulse leaves the populations unchanged allowing for the final symmetric ``reverse'' rotation; see Eq.~\eqref{eq:G3} for comparison.}
    \label{fig:zzz_ryd}
\end{figure}

\begin{figure}
    \centering
    \includegraphics{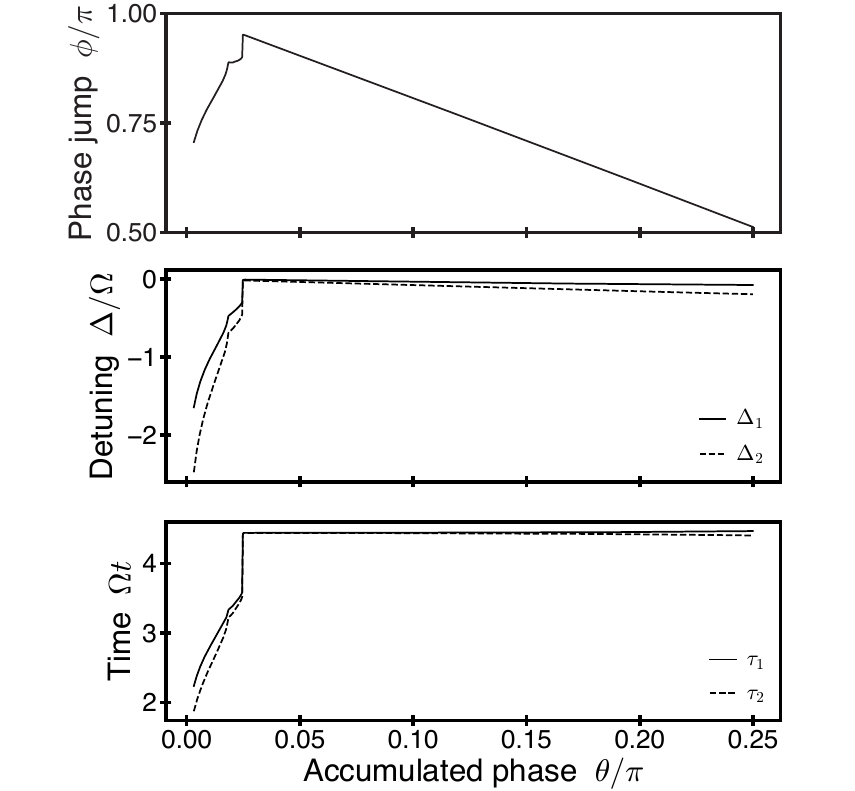}
    \caption{\textbf{Three-qubit gate: Pulse parameters.} Values of the phase jump, detunings, and timings for a given accumulated phase $\theta$. These plotted quantities define the $\mathcal{G}_3(\theta)$ pulse sequence through Eq.~\eqref{eq:G3}.}
    \label{fig:gate}
\end{figure}

Contrary to the original LP gate~\cite{Levine.2019}, the pulse time is not constrained by $\tau_i = 2\pi/\sqrt{2+\Delta_i^2}$, which would enforce that the $\ket{110}\to\ket{W_20}$ transition closes exactly after every pulse $R$. Amazingly, one can still find reasonable gates with this symmetry (up to 6 digits of precision), but for exact gates none of the manifold rotations close exactly. 
Instead, as shown in Fig.~\ref{fig:zzz_ryd}, the Rydberg populations of the $W_1$ and $W_3$ transitions are equal after every pulse. In fact, the population after the second pulse is the same as after the first, and the final rotation ``reverses'' the Rydberg state populated with the first pulse. In Fig.~\ref{fig:gate}, we plot pulse parameters that realize $\mathcal{G}_3(\theta)$ for $\theta$ between 0 and $\pi/4$. The discontinuity around 0.03 is caused by the necessity to perform an additional rotation on the Bloch sphere in order to acquire phases larger than the threshold value. Similar jump occurs again for larger angles but is outside of the range we plot. In general, optimal-control methods can be used to find more efficient gate implementations~\cite{Jandura.2022}.
\section{Variational ansatz cost functions}\label{app:Heff}
In order to find a good set of angles $\{\theta\}$ for variational state preparation or effective Hamiltonian engineering, we need to perform optimization within the family of circuits defined by Eq.~\eqref{eq:circ}. Given a reasonable initial guess, we can use gradient methods, such as gradient descent, to quickly find reasonable solutions. 
Here, we describe how to calculate the cost functions and their gradients in terms of the free-fermion state $\Gamma$ and the Hamiltonian matrix $A_{ij}$.

First, let us consider state preparation and the cost function,
\begin{equation*}
	Q_{\rm VSP} = -\abs{\braket{\psi_{\rm fin}|U_D(\{\theta\})|\psi_{\rm ini}}}^2,
\end{equation*}
based on the state overlap. The initial state $\ket{\psi_{\rm ini}}$ is captured by a matrix $\Gamma_{\rm ini}$ and the target state $\ket{\psi_{\rm fin}}$ is captured by $\Gamma_{\rm fin}$. The overlap between two pure states $\psi_1,\psi_2$ defined by $\Gamma_1$,$\Gamma_2$ is~\cite{Bravyi},
\begin{equation}
    \abs{\braket{\psi_1|\psi_2}}^2 = \pf[(\Gamma_1+\Gamma_2)/2],
\end{equation}
where $\pf[M]$ represents a Pfaffian of the matrix $M$. Therefore, the gradient is given by
\begin{align*}
    \partial_{\theta_i} Q_{\rm VSP} = -\frac{1}{2}\pf[(\Gamma_{\{\theta\}}&+\Gamma_{\rm fin})/2]\times\\
    &\tr[(\Gamma_{\{\theta\}}+\Gamma_{\rm fin})^{-1}\partial_{\theta_i}\Gamma_{\{\theta\}}],
\end{align*}
where we defined $\Gamma_{\{\theta\}} =\tilde{U}^\dagger(\{\theta\})\Gamma_{\rm ini}\tilde{U}(\{\theta\}) $ that describes the evolved state $U_D(\{\theta\})\ket{\psi_{\rm ini}}$. The  $\partial_{\theta_i}\Gamma_{\{\theta\}}$ gradient is given by
\begin{equation*}
    \partial_{\theta_i}\Gamma_{\{\theta\}} = \tilde{U}^\dagger(\{\theta\})\Gamma_{\rm ini}\partial_{\theta_i}\tilde{U}(\{\theta\})+\partial_{\theta_i}\tilde{U}^\dagger(\{\theta\})\Gamma_{\rm ini}\tilde{U}(\{\theta\}),
\end{equation*}
where $\tilde{U}(\theta)$ describes evolution in the Majorana picture; c.f. Eq.~\eqref{eq:majev}.
The circuit can be efficiently evaluated by pre-calculating the eigendecomposition of the ansatz matrices, e.g.,
\begin{align*}
    e^{\theta ZZ}= Qe^{\theta \Sigma} Q^T\quad {\rm for}\quad  ZZ=Q\Sigma Q^T,
\end{align*}
where $Q,\Sigma$ are pre-computed and $\Sigma$ is a real block-diagonal matrix with 2x2 blocks, which makes the exponentiation very fast. In these terms, the gradient of a single ansatz term is
\begin{align*}
    \partial_\theta e^{\theta ZZ}= Q\Sigma e^{\theta \Sigma} Q^T,
\end{align*}
which means that the only operations involved are matrix multiplications and efficient 2x2 exponentiations.

On the opeartor level, we use the squared Frobenius norm
\begin{equation*}
    Q_{H} = \norm{U_D(\{\theta\})-e^{-iH\tau}}_F^2,
\end{equation*}
where $\norm{A}_F^2=\tr[AA^\dagger]$. This describes the distance in the operator space between the desired evolution over time $\tau$ and the effective one realized by the trotterized circuit $U_D(\{\theta\})$. In the Majorana picture it is given by
\begin{equation*}
       Q_{H} = \norm{\tilde{U}(\{\theta\})-e^{-A\tau}}_F^2, 
\end{equation*}
cf. Eq.~\eqref{eq:majev}.
The gradient is given by
\begin{equation*}
    \partial_{\theta_i}Q_H = 2\tr[(\tilde{U}(\{\theta\})-e^{-A\tau})\times \partial_{\theta_i}\tilde{U}(\theta)],
\end{equation*}
which can be efficiently calculated by applying methods described above.

\section{Intialization for state preparation procedure}\label{app:prep}
The variational state preparation discussed in the main text transforms the toric code state into the non-abelian chiral phase. In this section, we summarize steps necessary for the preparation of that initial state, the toric code state, one the honeycomb lattice and discuss the consequences of the changing ground-state degeneracy across the transition from one topological phase to another.

\subsection{Projection to the no-vortex sector}\label{app:proj}

To prepare the desired topological phase, within our framework, we use the fine-tuned VSP circuits developed above. However, utilizing these methods requires that we are already in the correct vortex sector, since the variational circuit ansatz~\eqref{eq:circ} contains only terms that commute with the plaquette operators. This is desirable as it guarantees that once we establish the correct vortex configuration $\mathcal{W}$, it will not change. In this section, we discuss the two possible methods for projecting into the no-vortex sector.

Consider the state $\ket{\psi_0}=\ket{0}^N$ where all spins are initially pointing down, which is the ground state of $H_Z$ for $J_z <0$. Now, we want to project the state into the desired symmetry sector $\ket{\psi_{\rm ini}}=\mathcal{P}\ket{\psi_0}$, 
\begin{equation}\label{eq:proj}
    \ket{\psi_{\rm ini}}=\prod_{p=1}^{N/2-1}\left(\frac{1+W_p}{2}\right)\ket{\psi_0},
\end{equation} 
which creates the toric code state~\cite{Kitaev.2006}; see Appendix~\ref{app:tc} for the details of the mapping. To realize the projector $\mathcal{P}$, we utilize non-unitary operations (either dissipation, or measurement and conditional correction) that can prepare a long-range entangled state faster than local unitary operations. This method has been used to prepare the toric code state in recent experiment~\cite{Bluvstein.2022} and has been generalized to the preparation of various complicated long-range entangled states~\cite{Ruben.2021,Tantivasadakarn.2021,Liu.2022}. We also consider a dissipative method for state preparation, which does not require mid-circuit measurement, that takes time linear in system size. 

We emphasize, however, that these approaches cannot be straightforwardly used to prepare our final target state which has a nonzero Chern number. 
Indeed, we heavily rely on the fact that the projected state $\ket{\Psi}$ corresponds to the toric code state and the remaining $\mathbb{Z}_2$ vortices can be moved by applying products of Pauli strings, a particularly simple constant-depth operation; see Appendix~\ref{app:tc} for details.
\begin{figure}
    \centering
    \includegraphics{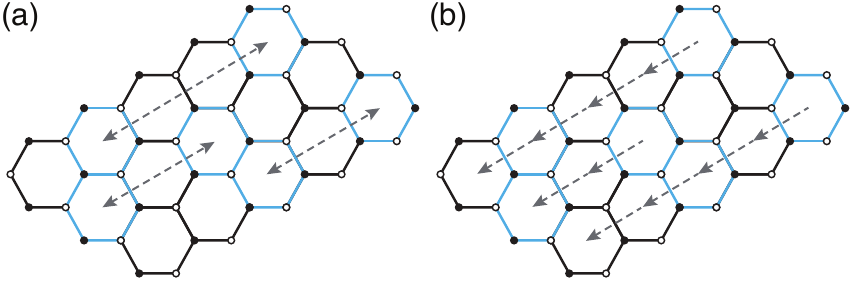}
    \caption{\textbf{Projection to the vortex-free sector.} The starting point of the variational state preparation is the toric-code state that can be obtained by removing the vortices from the initial product state. (a) Projective measurement of the $W_p$ plaquette operators and subsequent pairing of vortices in $O(1)$ time. The grey arrows symbolically indicate the transport of atoms, which is implemented by applying the $Z$ Pauli strings depicted in Fig.~\ref{fig:tc_map}. (b) Conditional transport of vortices to a single plaquette in every row, which effectively results in cooling the system to the no-vortex sector in $O(\sqrt{N})$ time.}
    \label{fig:proj}
\end{figure}

We propose two schemes to experimentally perform the projection $\mathcal{P}$ from Eq.~\eqref{eq:proj} . 
In both cases, we start by preparing the system in a product state $\ket{0}$ on each site.
In the first approach [Fig.~\ref{fig:proj}(a)], we measure all plaquette operators $W_p$ using one ancilla qubit per plaquette. 
This is similar to the procedure performed experimentally in Ref.~\cite{Bluvstein.2022} although here the operators are supported on 6 sites (which requires 6 control-Z gates per ancilla). Measuring the ancillas projects the system into a specific vortex sector, with vortices corresponding to $W_p = -1$, with the additional constraint of even number of vortices in every row. Then, we remove the vortices by pairing them up within each row according to a prescription given by, e.g., the minimum weight perfect matching algorithm~\cite{Edmonds.1965}. The vortices are moved between plaquettes within a row by applying Pauli $Z$ strings connecting them. This procedure prepares a no-vortex state in constant time $O(1)$ but it requires mid-circuit measurement and feed-forward capabilities. However, these particular capabilities may be inconvenient in near-term devices.

The second approach removes the measurement requirement but instead takes $O(\sqrt{N})$ time [Fig.~\ref{fig:proj}(b)]. Here, we again use ancillas to encode the plaquette expectation values but instead of projectively measuring the ancillas, we conditionally move the vortices by applying Pauli strings only if a vortex is present at the initial plaquette. Conditioning the movement in this way prevents the proliferation of vortices and can be implemented with a $CCZ$ gate. We choose a single plaquette in every row (left-most ones in Fig.~\ref{fig:proj}) as a \emph{sink} and direct the vortex flow towards that site where eventually all vortices get annihilated. This process is a form of many-body cooling and the energy is removed by resetting the ancillas after each round of transport. The time complexity of this procedure is proportional to the largest distance possible and thus scales with the linear system size $O(\sqrt{N})$.

\begin{figure}
    \centering
    \includegraphics[width=\linewidth]{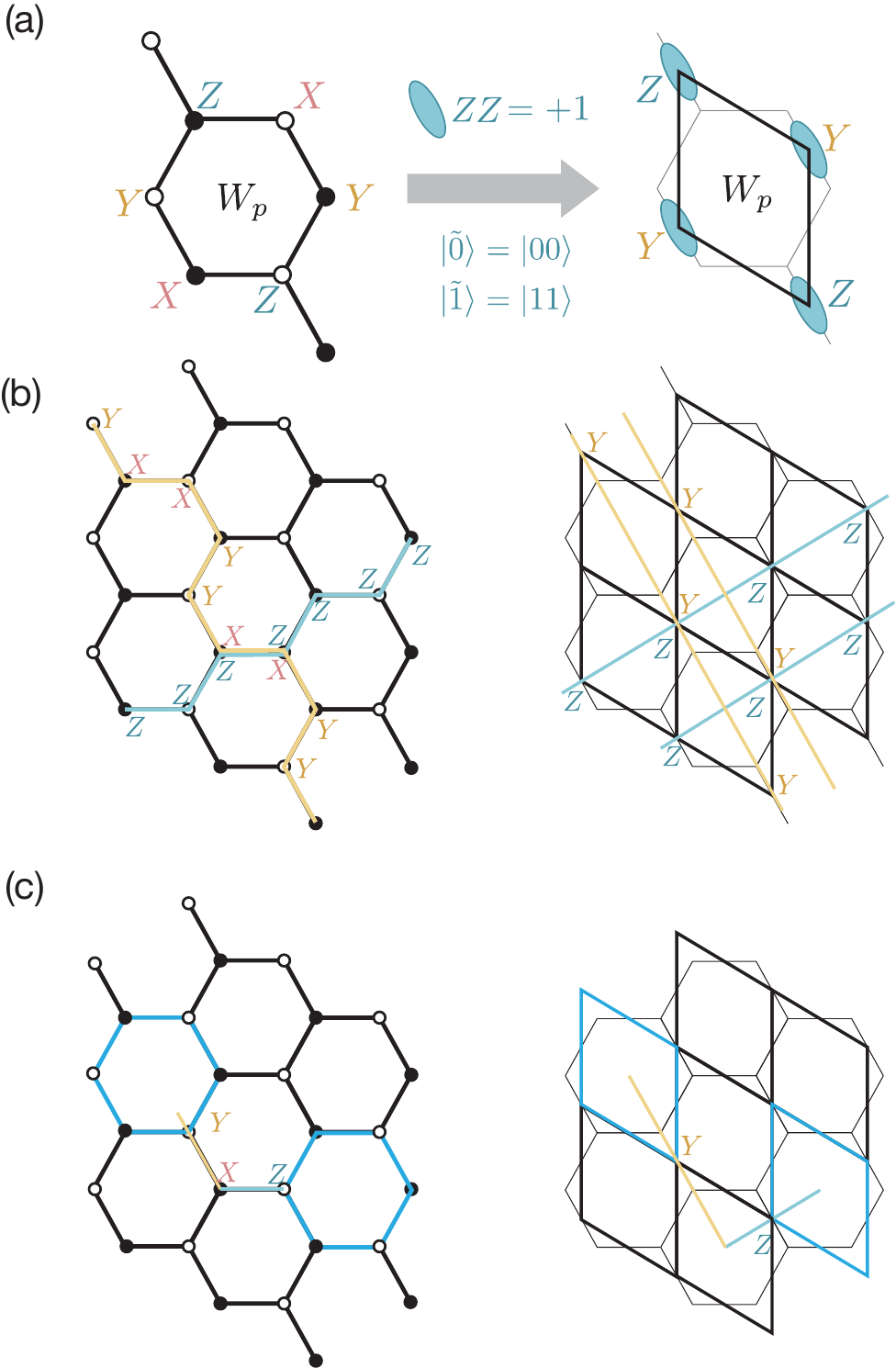}
    \caption{\textbf{Mapping between Kitaev honeycomb and toric code.} (a) By projecting all $Z$-links into the $ZZ=+1$ eigenspace, each link can be modelled by an effective qubit with two states, $\ket{\tilde{0}}=\ket{00}$ and $\ket{\tilde{1}}=\ket{11}$. Therefore, under this mapping, the six-body plaquette operator $W_p$ becomes a four-body plaquette, corresponding to the stabilizers of the Wen-plaquette model. This is also equivalent to the toric code model under single-site rotations. (b) The Kitaev honeycomb has two non-contracible loops which commute with all link-operators and all $W_p$, and hence do not change during time-evolution. Each of these operators map onto a pair of logical operators in the toric code. (c) Partial logical operators (open strings) of the Wen-plaquette model can be used to construct string operators in the honeycomb model which anti-commute with vortices living at their endpoints. A string connecting two vortices in the same row will consist solely of $Z$ operators.
    }
    \label{fig:tc_map}
\end{figure}

\subsection{Mapping to the toric code}\label{app:tc}
Here we briefly describe the mapping from the honeycomb model to the toric code in the $J_z\gg J_x,J_y$ limit, and see that the projection initializes the fixed-point of the toric code phase.

In this limit, we can approximate the low energy physics, as described in Ref.~\cite{Kitaev.2003}, by projecting into the $+1$ eigenspace of each $Z_i Z_j$ operator along a $Z$-type link.
This reduces the two qubits along a link to one effective qubit, spanned by two states $\ket{00}$ and $\ket{11}$.
After projection, the six-body $W_p$ plaquette operator becomes an effective four-body plaquette operator, acting as $YZYZ$ [see Fig.~\ref{fig:tc_map}(a)].
Thus, the $W_p=+1$ eigenspace corresponds to the groundspace of the Wen-plaquette model~\cite{WenPlaq.2003}, which is equivalent to the toric code model under a local unitary transformation.

Recall that during the initialization, all qubits are initialized in $\ket{0}^N$. Since all subsequent steps commute with $Z_iZ_j$ along $Z$-links, this ensures the final state has both $W_p=1$ for all $p$ but also $Z_iZ_j=+1$ for each $Z$-link. As such, the projection step prepares the fixed-point state at $J_Z > 0$, $J_X=0,J_Y=0$ which is the starting point for both adiabatic and variational state preparation of the non-abelian phase at $J_Z = J_X = J_Y$.
The initial state also introduces a constraint on the measured values of $W_p$. Namely, the product of $W_p$'s along a row of plaquettes must be even, since it is equivalent to the product of two non-trivial horizontal $Z$-loops. As a result, the number of $W_p=-1$ outcomes must be even in every row, and the corresponding anyons can be paired up by only applying $Z$ strings 
[see Fig.~\ref{fig:tc_map}(c)] as long as measurement errors are negligibly small.

\subsection{Degeneracy across the A-B transition}\label{app:transition}
The dimension of the ground-state manifold on a torus is dictated by the number of superselection sectors in the underlying topological quantum field theory. For the toric code phase this is 4, while for the non-abelian phase (Ising anyons) that number is 3. This means that one of the ground states is not adiabatically connected to the non-abelian theory and becomes a highly excited state. This state corresponds to the $(-1,-1)$ configuration of the two non-trivial loops on the torus, which are additional symmetries of the system with periodic boundaries, and are related to the logical operators in the toric code phase~\cite{Kells.2009} [Fig.~\ref{fig:tc_map}(b)]. By choosing an appropriate initial state
$\ket{\Psi}=\ket{0}^N$, we ensure that the system is orthogonal to this undesired state since $\ket{\Psi}$ is an eigenstate of the horizontal $Z$-loop with the $+1$ eigenvalue and all subsequent operations commute with the $Z$-loops.

\section{Readout fidelity with noisy preparation}\label{app:noisyvsp}
\begin{figure}
    \centering
    \includegraphics{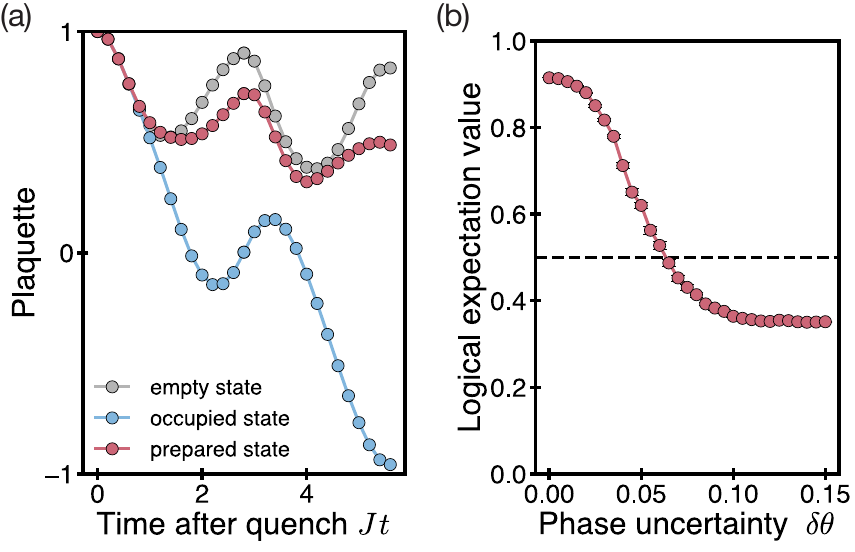}
    \caption{\textbf{Noisy state preparation of logical states.}  (a) Dynamics of the plaquette expectation value during the readout quench. (b) Logical expectation value of $(1+W_p)/2$ as a function of phase noise. This can be interpreted as the probability of identifying the state as $\ket{0}_L$. For $\delta\theta$\,$>$\,0.06, where the success probability falls below $50\%$, the state cannot be reliably identified.}
    \label{fig:sup_noisy}
\end{figure} 
The variational state preparation can be quantified in terms of the many-body overlap with the target state $Q_{\rm VSP}$. However, despite being in principle measurable~\cite{Bluvstein.2022} with interference techniques, this value has little utility for quantifying the quality of state preparation in an experiment. In this section, we study the performance of VSP in the presence of control noise, i.e., all phases $\{\theta\}$ in the preparation circuit are modified by a phase error from a uniform distribution $[-\delta\theta,\delta\theta]$. This error model captures, for example, the uncertainty in the global pulse parameters such as the pulse time or phase. This approach does not incorporate effects that violate translational invariance such as field inhomogeneity. The simulation was performed on an $L\times L$ lattice for $L=10$ with a circuit $U_{D=61}(\{\theta+\delta\theta\})$ and we averaged over a 100 noise realizations.

In order to quantify the robustness of state preparation, we prepare a $\ket{0}_L$ logical state in the two-dimensional logical subspace, similar to the fusion and braiding experiments. Then, we perform an ideal (noise-free) readout using the procedure described in Sec.~\ref{sec:readout}. In Fig.~\ref{fig:sup_noisy}(a), we show the readout quench dynamics for both reference logical states (empty/occupied) as well as the state prepared using a noisy circuit with $\delta\theta\approx0.03$. We find that the noise deteriorates the signal and effectively reduces the expectation value of the logical-state operator $(1+W_p)/2$. In Fig.~\ref{fig:sup_noisy}(b), we plot the expectation value of that operator as a function of the noise strength $\delta\theta$. The logical expectation value can be interpreted as a probability of measuring the target logical state; this value falls below $50\%$ around $\delta\theta\approx0.06$, which signifies that for noise above that value the logical state cannot be reliably prepared. We note that the expectation value saturates at a value $<0.5$, which signifies that for a noisy circuit the excited state is prepared with a higher probability than the ground state.

\section{Chiral edge modes}\label{app:chiral}
\begin{figure}[t]
    \centering
    \includegraphics[width=\linewidth]{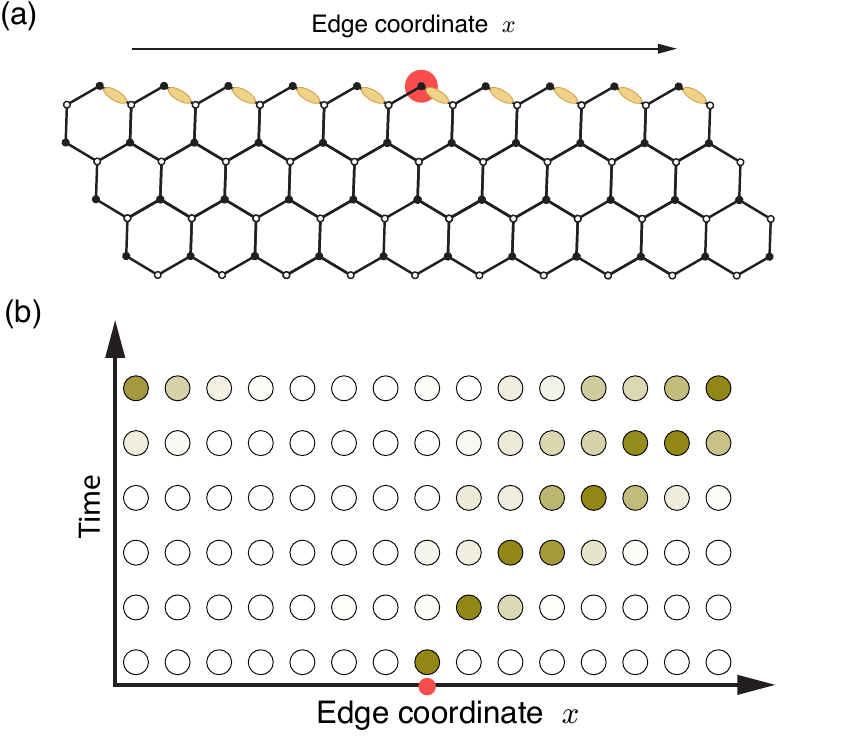}
    \caption{\textbf{Chiral edge modes.} The simplest experiment probing properties of the Kitaev $B$ phase is the observation of chiral edge states. Their presence manifests in a uni-directional transport after a single-qubit quench (red dot). (a) The system with a cylindrical topology: the edge $x$ direction is periodic while the other one has open boundary conditions. Measured operators are denoted with yellow links. (b) After the initial quench, the two-body edge correlations experience a wave-packet-like propagation due to chiral edge modes. The direction is dictated by the sign of the time-reversal symmetry breaking term.}
    \label{fig:sup_ce}
\end{figure}
The simplest physical experiment that can be performed using the tools introduced in this work is the observation of the chiral edge states. The Kitaev $B$ phase has a non-zero Chern number in the gapped bulk and so the bulk-edge correspondence~\cite{Hasan} guarantees gapless chiral edge modes. 

This means that transport at the edge has a preferred direction. For this purpose we will introduce mixed boundary conditions [Fig.~\ref{fig:sup_ce}(a)]: In one direction (the $x$ coordinate) we preserve the periodic boundaries, but in the other we use open boundary conditions; effectively, the system lives on a cylinder. The Hamiltonian in this geometry is realized by not applying gates to the links across the open boundary.

The edge modes are gapless and hence the ground state is difficult to prepare. However, for the purpose of observing the chiral properties of the system we can circumvent this issue. Instead, we prepare the gapped bulk phase on a torus and subsequently quench with the cylinder Hamiltonian. In a very short time, the edge modes will thermalize but the chiral response to subsequent perturbations will be preserved.

In experiment, it is straightforward to probe expectation values of Pauli operators and their higher moments---correlation functions. The $Y$ magnetization has been shown to exhibit a unidirectional transport~\cite{Mizoguchi.2020}, but for our purposes we propose to monitor edge correlators $\braket{Y_{2x}Y_{2x+1}}$ which naturally occur in the Hamiltonian and any perturbation that anticommutes with such terms should induce non-trivial dynamics. Here we perturb the system with a $Z$ operator ($\pi$-pulse) on a single site at the edge [Fig.~\ref{fig:sup_ce}(a)] and monitor the two-point spin correlation function. 

In Fig.~\ref{fig:sup_ce}(b), we show the time evolution of the normalized signal (the background value is subtracted) for a $30\times30$ lattice. Initially, the only affected observable is the one that includes the quenched site. This perturbation propagates to the right confirming the chiral nature of the edge. The wave packet slowly disperses and eventually loops around the boundary. Besides observing chiral edge modes in a digital simulation, this simple experiment can be used to verify the state preparation procedure as well as the effective Hamiltonian evolution.

\section{Magnetic fields and fermionic Gaussian states}\label{app:fgs}
In the presence of a magnetic field, the Kitaev Hamiltonian is no longer solvable. However, we were able to capture main features of the short-time dynamics using a variational ansatz for the wavefunction that is compatible with the Majorana solution of the model. 

In the original solution (without magnetic fields), the Hamiltonian is effectively quadratic since the bond Majorana pairs $u_{ij} = ib_i^{\alpha_{ij}}b_j^{\alpha_{ij}}$ are conserved and treated as clasical variables $[u_{ij},H]=0$; c.f. Eq.~\eqref{eq:Hmaj} and the discussion below. In terms of Majorana operators, a local $z$ field is expressed as $Z_j = ib_j^zc_j$ for the field at site $j$ and it does not commute with $u_{ij}$; this is because the magnetic field flips the plaquette operators $W_p$ and couples different vortex sectors. However, the field is local so it anticomutes only with the $\hat{u}_{jj'}$ on the corresponding $Z_j Z_{j'}$ bond but commutes with all others; thus, all but one bond variables remain conserved. We can capture this effect by breaking up the bond and working with the extended set of Majoranas $\{c_{1,...,N},b_j^z,b_{j'}^z\}$. The resulting Hamiltonian with the $-h Z_j$ magnetic field is
\begin{equation}\label{eq:fgsH}
    H[h] = H' - Jb_j^zb_{j'}^zc_jc_{j'} -K\sum_{k}u_{j'k} \,b_j b_{j'} c_jc_k -ih\,b_j^zc_j,
\end{equation}
where $H'$ corresponds to Eq.~\eqref{eq:Hmaj} with the bond $Z_j Z_{j'}$ removed and the $\sum_k$ term symbolically represents the four three-body terms that contain the ${j,j'}$ bond. The Hamiltonian in Eq.~\eqref{eq:fgsH} is no longer quadratic but we can make the calculation tractable and capture the main features of the new system by applying an appropriate wavefunction ansatz. In this case, we chose the fermionic Gaussian states (FGS) which have celebrated considerable success in simulating complex many-body systems~\cite{Shi.2018,Ashida.2018,Dolgirev.2020}

Now, we briefly describe the basic ideas behind the FGS approach while a detailed description, with non-Gaussian generalizations, can be found in Ref.~\cite{Shi.2018}. The variational ansatz is $\ket{\psi}=e^{-ic^TMc}\ket{{\rm vac.}}$ and the state is fully determined by the covariance matrix
\begin{equation}
    \Gamma_{ij} = \frac{i}{2}\braket{[c_i,c_j]},
\end{equation}
and thus satisfies Wick's theorem
\begin{equation}
    \braket{c_ic_jc_kc_l}=-(\Gamma_{ij}\Gamma_{kl}-\Gamma_{ik}\Gamma_{jl}+\Gamma_{il}\Gamma_{jk}),
\end{equation}
which allows us to simplify expressions considerably. The equations of motions in real and imaginary time are given by
\begin{subequations}
\begin{align}
   d_t \Gamma &= [\mathcal{H},\Gamma]  ,\\
   d_\tau \Gamma &= -\Gamma - \Gamma \mathcal{H} \Gamma,
\end{align}
\end{subequations}
where $\mathcal{H} := \mathcal{H}[\Gamma]=4\delta\braket{H}_\Gamma/\delta\Gamma$ is the functional derivative of the expectation value of the Hamiltonian. In the special case of quadratic Hamiltonians $H=(i/4)\sum_{i,j}A_{ij}c_ic_j$, this formalism is exact and $\mathcal{H}_{ij}=A_{ij}$---this is the case for the Kitaev model without magnetic fields.

Applying the above formalism to Eq.~\eqref{eq:fgsH} enables approximate simulation of the quench dynamics governed by the Kitaev Hamiltonian with local magnetic fields. We benchmark this method against exact results on a small 30-qubit system [Fig.~\ref{fig:readout}], where it performs well, and use it to predict the readout fidelity in a larger system [Fig.~\ref{fig:4}]. 

\section{Effective two level system}\label{app:tls}
The key insight behind the readout procedure introduced in the main text is that the state with a localized fermion ($\ket{1}_L$) and the state without one ($\ket{0}_L$) react differently to quenching a local magnetic field. The magnetic field operator $Z_j$  acting on the site $j$ is represented in the Majorana description by
\begin{equation}
    Z_j = ib^z_j c_j,
\end{equation}
where $b^z_j$ is the ``bond Majorana'' (see Appendix~\ref{app:fgs}) that enters in the plaquette operators and $c_j$ is the ``matter Majorana`` that forms the fermionic spectrum of the system. Crucially, $b_j^z$ anticommutes with the plaquette operator (see Appendix~\ref{app:maj}) and $c_j$ commutes with it; thus, we can decompose the $Z_j$ operator into two parts, one acting in the plaquette space and the other in the matter space.

Consider a situation with two pairs of zero modes where each pair occupies adjacent plaquettes (but pairs are separated), as in the inset of Fig.~\ref{fig:readout}. Since the distance between the two modes is small (on the order of the lattice spacing), they are no longer degenerate and the first excited state consists of fermions localized on the bonds shared by the occupied plaquettes. When deconstructed into Majorana particles, we expect that state to have large overlap with $c_j$, which we can symbolically decompose into $c_j \approx a+a^\dagger$, where $j$ denotes the index of a site on the bond. Now, if we apply a local magnetic field $Z_j = ib^z_j c_j$ to the vertex $j$, the response of the system will depend on whether the localized-fermion state is occupied or not. In both cases, the quench couples to a state with two real vortices (flipped plaquettes). Importantly, the fermionic spectrum of the two-vortex sector with two flipped bonds (used to create Majorana modes) is identical to the vortex-free sector of the uniform-bond Hamiltonian (with no flipped bonds).

If the localized mode is occupied, the $a^\dagger$ component of $c_j$ will not contribute and the resulting state amounts to annihilating the matter fermion with $a$ and flipping the two neighboring plaquette operators with $b^z_j$ (creating two vortices). This state is in fact the ground state of the uniform vortex-free sector and we can approximate the dynamics as an effective two-level system with the Rabi frequency proportional to the magnetic field. Additionally, the vortex-free ground state has a slightly different energy from the initial state, which results in a non-zero detuning $\Delta$ in the two-level system. In total, we have a product state of two spin-1/2 states, one from each Majorana pair.

If the localized mode is not occupied, the $a$ operator has no effect and the system is coupled to a two-vortex state where $a^\dagger$ creates a localized fermion. However, there is no localized fermion in the spectrum of the two-vortex sector with a flipped bond which means that the system couples weakly to a continuum of momentum modes, with very small overlaps. This results in weak, damped oscillations in the plaquette expectation values.

To validate this two-level picture, we can extract the energy difference between the states and the state overlaps, and subsequently compare these parameters to the Rabi frequency and detuning extracted from numerics. We find that the two agree up to a few percent.

\section{Calibrating the readout procedure}\label{app:rcal}
\begin{figure}
    \centering
    \includegraphics{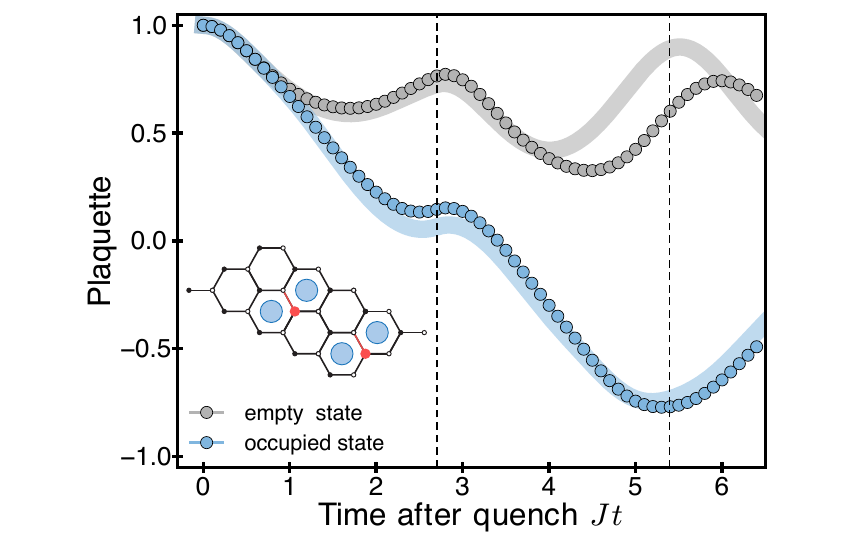}
    \caption{\textbf{Readout calibration.} Plaquette expectation value after quench of the occupied and empty state. The shaded lines correspond to the FGS time evolution with the full Hamiltonian while the circles are the result of exact simulation in the spin picture using engineered Floquet Hamiltonian (as would happen in the experiment). The dashed line marks the end of the first pulse. (inset) The 30-spin lattice configuration with periodic boundary conditions and the location of Majorana modes. The magnetic field acts on the sites marked with red dots.}
    \label{fig:readout}
\end{figure}
We propose a small-scale (30 qubit) experiment to both verify the state preparation and calibrate the readout procedure. The lattice consists of $3\times5$ plaquettes and the target phase has a state with two pairs of adjacent Majorana modes in its spectrum; see inset of Fig.~\ref{fig:readout}. This mode can be either occupied ($\ket{1}_L$) or empty ($\ket{0}_L$). 

Because the system is so small, we can easily target both $\ket{1}_L$ and $\ket{0}_L$ with variational state preparation, which removes the need for applying the effective Hamiltonian  thus significantly shortening the circuit depth. Then, we perform the magnetic field quench using the effective Hamiltonian and calibrate the readout procedure without implementing neither fusion nor braiding operations. The readout of the particle content is a local operation and we expect that the procedure calibrated in a small system will perform similarly well. In fact, the pulses applied in Figs.~\ref{fig:4} and~\ref{fig:readout} have exactly the same magnetic field and first-pulse timing.

In Fig.~\ref{fig:readout}, the readout procedure has been simulated exactly (in the original spin picture) to remove any errors due to the FGS approximation. The good agreement with the full Hamiltonian evolution in the FGS approximation (shaded lines) suggests that the FGS is a reasonable predictor for readout fidelities in larger systems.

\section{Lattice gauge theories with three-qubit gates}\label{app:lgt}
The goal is to simulate the (1+1)D lattice gauge theory (LGT) Hamiltonian $H=H_{\rm g}+H_{\rm m}+H_{\rm m-g}$,
\begin{align}
H_{\rm g} &= -f \sum_i L^2_{i,i+1},\nn\\
H_{\rm m} &= \mu\sum_i c^\dagger_i c_i,\nn\\
H_{\rm m-g} &= -(J/2)\sum_{i} (c_{i+1}^\dagger U_{i+1,i} c_i+{\rm h.c.}),
\end{align}
where the $c_i^\dagger$ ($c_i$) are the fermion creation (annihillation) operators satisfying the canonical anticommutation relations (CAR), $L^2_{i,i+1}$ represents the energy of the electric field on the $\{i,i+1\}$ bond, and $U_{i+1}$ is the gauge-matter-coupling operator. Note that there is no magnetic field energy since for one spatial dimension the only gauge-invariant combination of $U_l$s would be a loop around the whole system.

The most difficult step in making such theories amenable to digital spin simulations is the treatment of the fermionic CAR. Usually, the resulting spin Hamiltonian involves conditional evolution and additional degrees of freedom~\cite{VC.2005}, where the fermionic phases are tracked by ancillas. Fortunately, in (1+1)D the situation is much simpler: No conditional evolution or ancillary qubits are required.
This results in the typical spin Hamiltonian,
\begin{align}
H_J &= -J\sum_{i} \tau_{i,i+1}^x(X_i X_{i+1}+Y_iY_{i+1}),\nn\\
H_f &= -\lambda_E \sum_i \tau_{i,i+1}^z,\nn\\
H_m &= \mu\sum_i (-1)^i Z_i,
\end{align}
which corresponds to Eqs.~(1a)-(1c) in Ref.~\cite{Mildenberger.2022}. Notice that the matter-gauge coupling is now given by the $XXX$ and $XYY$ operators.

The most expensive term to simulate is the three-qubit matter-gauge coupling, which under general decomposition schemes requires around 20 two-qubit operations~\cite{Shende.2006} but in Ref.~\cite{Mildenberger.2022} the authors performed the entire single trotter step with just 8 two-qubit gates. Here we propose to use the native $\mathcal{G}_3$ gates to realize the same trotter step with 2 three-qubit gates combined with sublattice rotations. Moreover, Ref.~\cite{Mildenberger.2022} was limited to $\theta \approx \pi/4$ which severely limits their control over the trotter step and associated heating. While they argued (and supported their claim with numerical simulations) that this drive frequency is well within the regime of slow heating, the full flexibility of $\mathcal{G}_3(\theta)$ allows not only a more precise digital simulation with a smaller trotter error but also a comprehensive study of heating processes in LGTs.

Naively, we would require 4 gates ($XXX$ and $XYY$ for each of two neighbors) but a natural Floquet echo trick might reduce it to 2. Instead of performing $XXX$ and $XYY$ separately, we instead apply a $\pi/4$ rotation resulting in $X(X+Y)(X+Y)$. This is the gate we want up to the mixed terms $XYX$ and $XXY$. However, in the subsequent round we can do a $-\pi/4$ rotation resulting in $X(X-Y)(X-Y)$ such that, to first order, the mixed terms cancel out. This methods gives us a $4\times$ reduction in the circuit depth over Ref.~\cite{Mildenberger.2022} while retaining full control over the accumulated phase $\theta$.

\bibliography{bibliography.bib}
\end{document}